# Effect of Random Fiber Network and Fracture Toughness on the Onset of Cavitation in Soft Materials

Fuad Hasan[1], KAH Al Mahmud[1], Md Ishak Khan[1], Wonmo Kang[2], Ashfaq Adnan[3]

1. Ph.D. Student, Department of Mechanical and Aerospace Engineering, The University of Texas at Arlington.
2. Assistant Professor, School for Engineering of Matter, Transport and Energy, Arizona State University.
3. Professor, Department of Mechanical and Aerospace Engineering, The University of Texas at Arlington.

**Corresponding Author:**
Dr. Ashfaq Adnan,
Professor
Department of Mechanical and Aerospace Engineering
The University of Texas at Arlington
Woolf Hall, Room 315C, Arlington, TX 76019, USA.
T 817-272-2006; F 817-272-5010
aadnan@uta.edu

**Abstract**

Dynamic cavitation has the potential to cause tissue damage. Therefore, it is crucial to identify the critical condition of cavitation initiation in soft materials (e.g., tissue). Experimental and theoretical observations have agreed that the onset of cavitation in soft materials requires higher tensile pressure than pure water. The extra tensile pressure is required since the cavitating bubble needs to overcome the elastic energy in soft materials. In this manuscript, we have proposed that i) the critical bubble nuclei size is comparable to the soft materials' microscopic length scale, and ii) the threshold tensile pressure is required to activate the bubble nuclei (e.g., nucleation pressure) and overcome the elastic energy (e.g., extra tensile pressure). We developed two models to study and quantify the extra tensile pressure. In the first approach, we proposed a strain-energy based random fiber network (RFN) failure criteria in which interaction between the cavitating bubble and RFN is considered. Gelatin samples are prepared for different concentrations, and SEM images are used to study the microstructural properties of the RFN. A unit-cell model is introduced to evaluate the geometrical and mechanical properties of the RFN. The network strain energy formulation is then coupled with the bubble growth, and the critical condition is set as the fibers' ultimate failure strain. We considered soft materials as homogenous hyper-elastic Ogden material, and fracture-based failure criteria are proposed in the second approach. The critical energy release rate (e.g., fracture toughness) is considered for quantifying the extra tensile pressure. Both the models are then compared with the existing cavitation onset criteria of rubber-like materials. The validation is done with the experimental results of threshold tensile pressure for different gelatin concentrations. We have found that due to the large distribution of the pore size in the network, the nucleation pressure is similar to water nucleation pressure. Both models can moderately predict the extra tensile pressure within the intermediate range of gelatin concentrations (3-7% [w/v]). For low concentration (~1%), the network's non-affinity plays a significant role and must be incorporated. On the other hand, for higher concentrations (~10%), the entropic deformation dominates, and strain energy formulation is not adequate.

**Keywords:** energy release rate (A), fracture toughness (A), microstructures (A), biological material (B), viscoelastic material (B), electron microscopy (C), cavitation.

**Graphical Abstract**

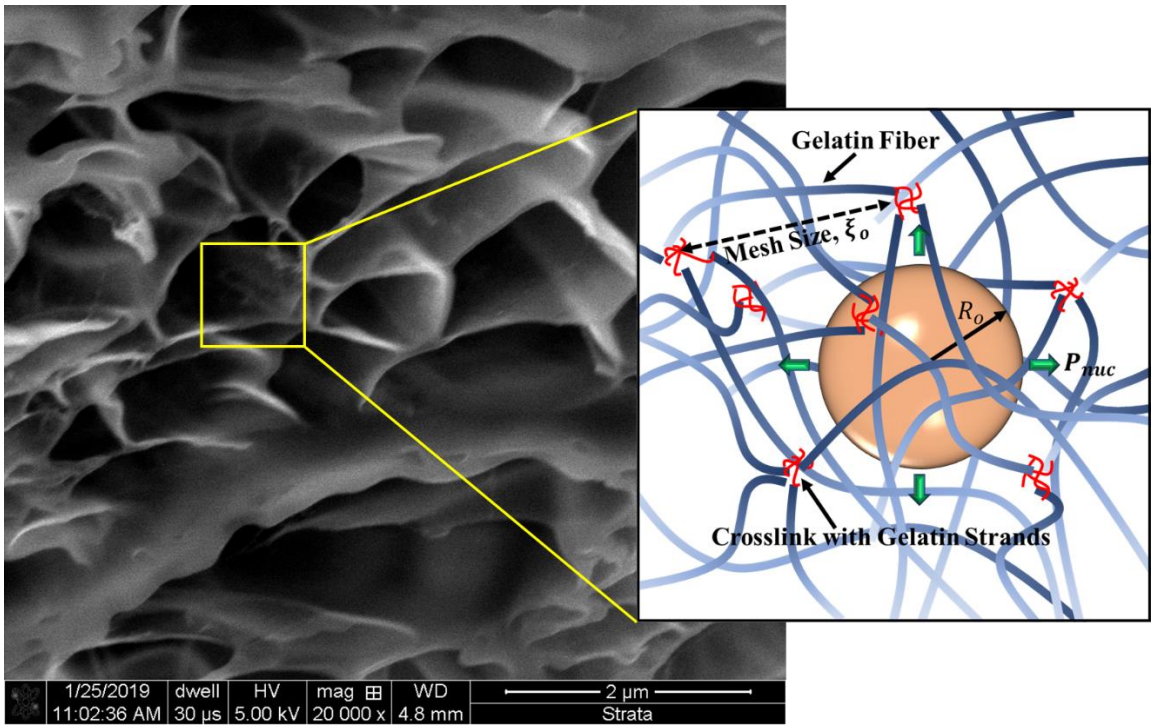

Graphical Abstract: Soft Materials, when subjected to tensile pressure, may experience cavitation. At the nucleation pressure, bubble nuclei start to grow within the soft materials' random fiber network and need to overcome the network's elastic energy before they can cavitate. The network fails via fibers' ultimate failure strain at the threshold tensile pressure, and the bubble grows to visible size.

1. **Introduction**

Cavitation is considered one of the main driving factors that can potentially cause soft tissue damage [1]. The phenomenon has been studied in the medical treatments (e.g., lithotripsy, drug delivery, etc.) applications from the biological perspective and recently has gained its fair share of attention since it has been linked to Traumatic Brain Injury (TBI) [2], [3]. This implication leads to the study of cavitation bubble dynamics in soft tissue like materials (e.g., gelatin hydrogel). The recent research on cavitation has intended towards the origin of nucleation theory and macroscopic behavior of cavitating bubbles in soft materials [4]. Kang et al. (2018) used a novel drop-tower system to impact the gelatin samples and characterized critical acceleration to cavitation inception for various gelatin gel concentrations. They suggested corresponding critical tensile pressure ($P_T$) required for the onset of cavitation in gelatin and showed that there is a 175% increase from the water to 1% [w/v] gelatin. Evidently, soft materials tend to withstand more tensile load before the inception of cavitation damage. Hence, critical tensile pressure seems to be one of the most important tissue damage study parameters. In this context, we should mention the work of Gaudron et al. (2015), who modified the well-known Rayleigh-Plesset equation of bubble dynamics for a nonlinear viscoelastic model to study the bubble dynamics in soft materials [5]. Their study on the stability of bubble nuclei showed that elasticity plays a significant role in stabilizing much smaller nuclei by modifying Blake's radius [6]. It can be argued that higher activation energy is required than water for cavitation to happen in the soft materials having a random fiber network (RFN), which gives their elastic properties. Mahmud et al. (2020) observed the interactions of collagen RFN with bubble growth in their molecular simulation study and agreed that higher tensile pressure is required [7]. These three studies mentioned above dictated our motivation for this manuscript to systematically study the threshold tensile pressure of cavitation onset in soft material. We have postulated that i) due to the presence of the random fiber network of gelatin gel, the critical flaw size (e.g., bubble nuclei of radius $R_0$) is comparable to the pore size of the gel network and ii) the threshold tensile pressure ($P_T$) is required for two consecutive works done by the bubble. First, a portion of the tensile energy is spent on activating the bubble nuclei and then rest of the energy is spent on overcoming the surface energy and the elastic energy imparted by the gel system for bubble to cavitate.

Recent studies on cavitation in soft biomaterials are based on the numerous pioneering works on cavitation in hyperelastic elastomers (e.g., rubber) [8]–[11]. We have found that one of the most frequently used cavitation onset criteria for visible bubbles to appear in a rubber block subjected to far-field triaxial tension is when cavity pressure ($P_c$) reach a critical value of $5G/2$ ($G$ is the shear modulus for small strain) [12]–[14]. This onset criterion is based on the incompressible Neo-Hookean material model and applicable to nuclei size ranging from $0.5\ \mu m$ to $1 mm$ (the range is only for rubber like material with higher elastic modulus and fracture toughness than gelatin hydrogel) [15]. In the "cautionary tale" by Gent, he discussed the limitation of this onset criterion and suggested, i) for smaller nuclei surface energy might dominate the cavity growth, ii) at large deformation, real rubber-like materials do not follow the simple kinetic theory of rubber-like elasticity and iii) a fracture based approach based on Griffith's fracture criterion [16] might explain the anomaly for smaller nuclei [17]. Indeed, Williams and Schapery (1965) considered the energy required for the cavities to fracture and showed that the required cavity pressure was as large as $9G$ for $0.5\ \mu m$ nuclei and even more for much smaller nuclei [18]. In the recent trend, more realistic material models have been used with "energy limiter" to study the cavitation in rubber [19], [20]. At the same time, Lopez-Pamies argued that cavitation needs to be studied as dynamic deformation where both inertia and viscosity play a significant role in bubble dynamics along with the material nonlinearity [21]–[23]. However, he discussed the poor agreement between theoretical and experimental observations and suggested that rubber fracture's microscopic mechanism needs to be incorporated for further study [24]. Indeed, all the references mentioned above did not consider the heterogeneity posed by the rubber material's microstructure (e.g., the interplay between RFN and bubble). The argument raises the question of the application of cavitation onset criterion ($9G \geq P_c \geq 5G/2$) for gelatin gel, or in general, for any biomaterials because of the inherent microstructure of hydrogel and rubber are quite different. The significant difference between rubber and biomaterials is that the microstructure of rubber is made of RFN of "flexible fibers/chains" having "entropic deformation" [25] whereas biomaterials have network of "semi-flexible fibers" having "enthalpy dominant deformation" [26]. Moreover, studies involved in cavitation in rubber considered presence of "vacuous cavity" while homogeneous nucleation theory of cavitation considers the presence of nuclei as a spontaneous phase change phenomenon between the liquid phase (water) and cavity filled with the vapor and noncondensable gas under tension [27]–[31]. Therefore, considering gel system as a biphasic material where RFN is the solid phase and solvent (e.g., water) is the liquid phase, we can assume that the nucleation process is same as in the water and solid phase (i.e., RFN) only interacts while the bubble grows. This assumption is the base of our second postulate mentioned in the previous paragraph. This approach simplifies the analysis by separating the contribution of tensile load to two consecutive events; first, to provide enough activation energy to nucleate a bubble and second, to overcome elastic and surface energy for bubble to grow.

However, the elastic contribution of the gelatin network on bubble growth is complicated to quantify. There are mainly three different approaches to study the mechanics of the RFN [32], [33]: Single fiber mechanics [34], [35], unit cell modeling [36]–[39], and construction of the 3-dimensional (3D) network [40]–[50]. The structure of the gelatin network itself is very complex [51]–[53] and depends on many factors, e.g., source of gelatin (bovine, porcine, etc.), manufacturing process, and sample preparation [54]–[58]. Moreover, 3D network generation requires substantial computational effort; therefore, we have adopted the "unit cell modeling" approach. In this approach, one needs to propose a unit cell having a finite number of fibers, followed by establishing a relationship between the fiber stretching ($\lambda_f$) and the unit cell stretching ($\lambda_G$). Then utilizing the experimental stress-strain (stretching) data we can use a curve fitting method to find the relevant properties of the fibers. We have seen successful implementation of the unit cell modeling approach in rubber elasticity with "flexible fibers/chains". A 3-chain [59], [60], 4-chain [61], [62] and 8-chain [63], [64] models are the most frequently adopted unit cell models. Recently 8-chain model of Arruda and Boyce (1993) has been modified for "semi-flexible fibers" and used to describe mechanics of fibrin network and mussel byssal threads [65], [66]. Cryogenic-temperature scanning electron microscopy (cryo-SEM) observation shows that gelatin fibers form a d-periodic hierarchical structure similar to collagen fibril and connected via triple-helical gelatin strand at the crosslink [67]. Due to the similarity of collagen fibril and gelatin fiber we have used the unit cell proposed by Susilo et al. (2010) who studied the micromechanics of collagen based extra-cellular matrix (ECM). Experimental stress-strain data up to the gel failure is then used to find the fiber level ultimate failure stretch ($\lambda_f^u$) [68], [69].

The present work investigates the cavitation threshold tensile pressure for the gelatin gel system. In the first approach, the role of RFN on bubble growth is evaluated. The failure of the individual fiber is considered as the limiting criterion for cavitation onset. The strain energy gained by the RFN before the failure is used to calculate the tensile pressure. In the second approach, the critical energy release rate ($G_c$) is used as a fracture criterion considering gel as a homogeneous medium, and a comparison is drawn between the two approaches. The present manuscript is organized as follows. First, we have discussed in detail the justification of our assumptions. The length scale involved in the gel system for bubble nucleation is presented, and threshold tensile pressure ($P_T$) is defined. In the next section, mesh size (e.g., crosslink to crosslink fiber distant, $\xi_0$) is computed using the linear shear modulus ($G$) and critical flaw size ($R_0$) is proposed based on the pore size. A unit cell model is then used to quantify the fiber properties, and the bubble growth mechanism is introduced to the unit cell failure. Lastly, a fracture criterion is formulated for the hyperelastic Ogden material model [70], followed by results and discussion sections.

## 2. Cavitation Threshold Tensile Pressure ($P_T$) for Gelatin

The theory of homogeneous nucleation states that the random thermal motion of atoms spontaneously creates energetic particles that leave the liquid phase and vaporize, thus form nucleation sites [31]. Church (2002) showed that the required energy for the nucleation increases with no bound at atmospheric pressure condition ($P_{atm}$) and nuclei immediately collapse. A liquid subjected to negative (tensile) pressure, however, becomes "metastable," meaning the required energy is finite for spontaneous nucleation to occur, and it depends on the strength of the tensile pressure, surface tension, and temperature.

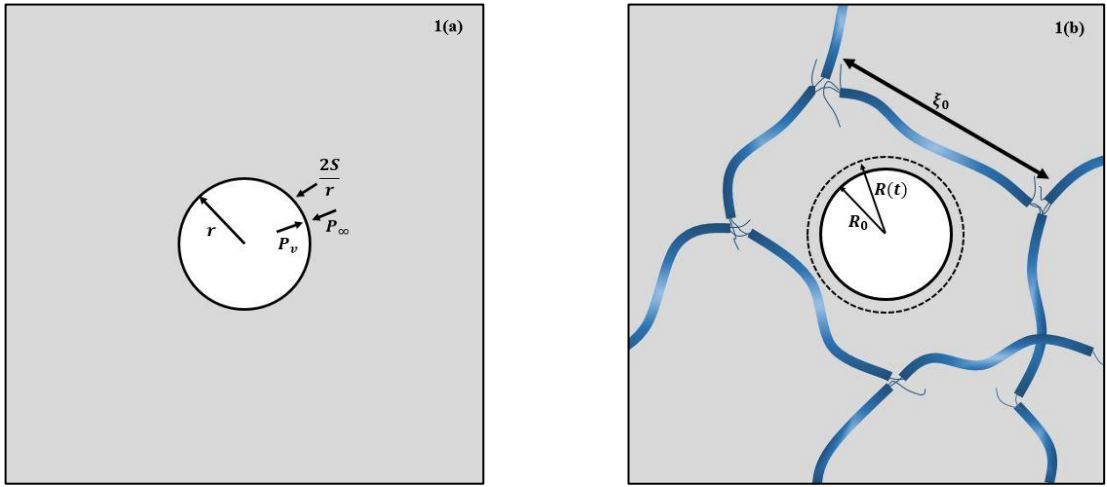

Fig. 1. Bubble nucleus in pure liquid (a) and in a gel (b). In pure liquid, the bubble pressure is in mechanical equilibrium with far-field pressure and surface tension (a). When a bubble starts to grow in the fiber network's presence (in gel), it needs to overcome the resistance hence requiring extra energy to cavitate (b).

As described by Herbertz (1988), the nucleation energy ($W_{nuc}$) required to form a nucleus under tension consists of three work terms [71]: i) $W_c$, the work to create the cavity under far-field tensile pressure ($P_\infty$), ii) $W_i$, the work needed to establish the interface having surface tension ($S$), and iii) $W_v$, the energy attained by the formation of the vapor phase at vapor pressure ($P_v$) (see Fig. 1a),

$$W_{nuc} = W_c + W_i - W_v$$
$$= 4\pi r^2 S + \frac{4}{3}\pi r^3 (P_\infty - P_v) \qquad (1)$$

Eq. (1) implies that for a given tensile pressure ($P_\infty = P_{nuc}$) there exists maximum nucleation energy, $W_{nuc,max}$ (at $dW_{nuc}/dr = 0$), which is required to form the bubble of the critical radius, $r = R_0$. Fig. 2 shows the energy necessary for different far-field pressure. For any positive (compressive) pressure, the required energy is unbounded. However, for negative pressure, if the available energy is less than that of maximum energy ($W_{nuc,max}$), then the bubble will have a smaller radius than the critical radius ($R_0$) and will eventually collapse following the left-hand side of the curve. On the other hand, the bubble will grow and cavitate if the radius is bigger than the critical value. For a simple homogenous liquid, the tensile pressure will suffice to cavitate, given that the condition is met as described above. The tensile strength of the liquid is defined as $\Delta P = (P_v - P_{nuc})$. From Eq. (1), the critical energy of nucleation can be found from the tensile strength of the liquid considering the critical tensile pressure ($P_\infty = P_{nuc}$) and saturation pressure of the bubble content,

$$W_{nuc,max} = W_{cric} = 4\pi R_o{}^2 S + \frac{4}{3}\pi R_o{}^3 \Delta P \tag{2}$$

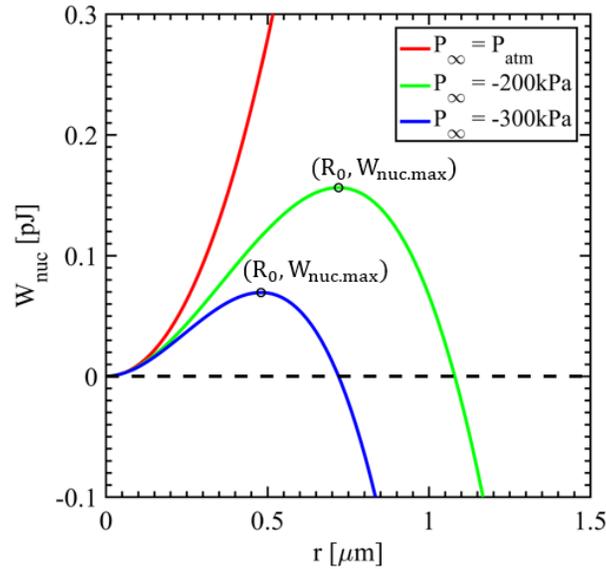

Fig. 2. Nucleation energy as a function of nuclei radius for a given pressure (Eq. (1)). The critical energy, critical radius, and nucleation pressure are indicated at each curve's peak (Eq. (2)).

Now in the presence of the RFN, the required tensile pressure will be significantly higher than that of pure liquid, considering the same bubble nucleus to grow from the critical radius ($R_0$) (Fig. 1b). Therefore, the threshold tensile pressure for gelatin is defined as,

$$P_T = P_{nuc} + \Delta P_T \tag{3}$$

In the above equation, extra tensile pressure ($\Delta P_T$) arises due to the bubble growth in the random fiber network and surface energy [4]. Fig. 1b depicts this scenario where gray background and blue fibers represent bulk liquid and fiber networks, respectively. To the best of our knowledge, until now, only Kang et al. (2018) reported the cavitation threshold tensile pressure ($P_T$) for different concentrations ($C\%$ [$w/v$]) of gelatin gel (Fig. 3). In their work, "Knox gelatin" is solved with water in weight to volume ratio to prepare 1%, 2.5%, 5%, and 7.5% gelatin gels. In Fig. 3, the green dotted line indicates a 175% increase in threshold tension from water to 1% gelatin. The Blue dotted line is fitted with a nonlinear least square method ($R^2 = 0.93$) to show the increasing trend with gelatin concentration. The ultimate goal of this manuscript is to quantify $P_T$ for the different concentrations of gelatin. From Eq. (3), it is apparent that we need to identify the contribution of $P_{nuc}$ and $\Delta P_T$, respectively.

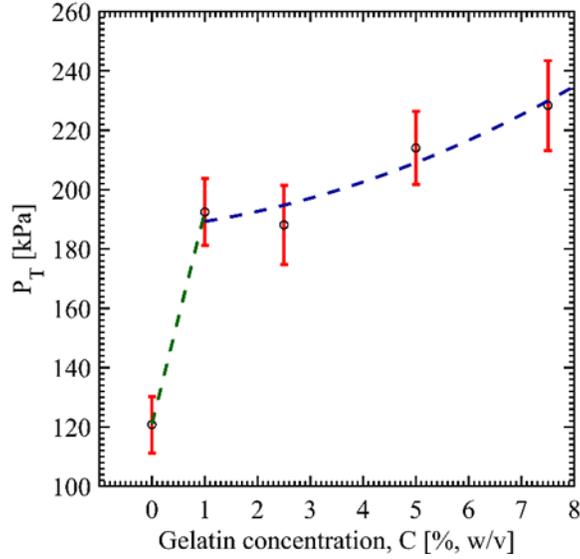

Fig. 3. Experimental observation of the threshold tensile pressure for different concentrations of gelatin gel [4]. C=0 corresponds to pure water. The vertical red error bar indicates the standard deviation.

*2.1 Critical Radius and Nucleation Pressure*

Carey (1992) considered a system where a bubble of a radius $R_0$, is in a bulk liquid under tension ($P_{nuc}$) [72]. His formulation satisfies both mechanical and thermodynamic equilibrium for that system to be stable. At the thermodynamic equilibrium, both the vapor and the liquid phase will have the same chemical potential ($\mu_i$),

$$\mu_l = \mu_v \tag{4}$$

Where subscripts $l$ and $v$ are for the liquid and vapor phase, respectively. Now, We can use the Gibbs-Duhem relation ($d\mu_i = -sdT + vdP$) for both vapor and liquid phase, where $\mu_i$ is the chemical potential of the $i^{th}$ phase, $s$ is specific entropy, $T$ is temperature, $v_i$ is the specific volume of the $i^{th}$ phase and $P$ is pressure. Integrating the Gibbs-Duhem relation from saturation condition to any arbitrary pressure at a constant temperature ($T_\infty$),

$$\mu_i - \mu_{i_{sat}} = \int_{P_{sat}(T_\infty)}^{P} v_i dP \tag{5}$$

Considering bubble content as an ideal gas ($v_v = R_v T_\infty / P$), where $R_v$ is a specific gas constant for vapor, the chemical potential for the vapor phase from the saturation condition to bubble vapor pressure ($P_v$) using Eq. (5) is,

$$\mu_v = \mu_{v_{sat}} + R_v T_\infty \ln \left[\frac{P_v}{P_{sat}(T_\infty)}\right] \tag{6}$$

Similarly, for the incompressible liquid phase ($v_l = constant$) the chemical potential from the saturation condition to arbitrary tensile pressure ($P_{nuc}$) using Eq. (5) is,

$$\mu_l = \mu_{l_{sat}} + v_l [P_{nuc} - P_{sat}(T_\infty)] \tag{7}$$

Now substituting Eq. (6) and Eq. (7) into Eq. (4) and considering thermodynamic equilibrium at the saturation condition as well ($\mu_{l_{sat}} = \mu_{v_{sat}}$) we get,

$$P_v = P_{sat}(T_\infty) \exp\left\{\frac{v_l[P_{nuc} - P_{sat}(T_\infty)]}{R_v T_\infty}\right\} \tag{8}$$

The above equation is based on the thermodynamic equilibrium. Considering bubble only contains saturated vapor and no non-condensable gas, then mechanical equilibrium at the formation of a bubble in the pure liquid requires,

$$P_v = P_{nuc} + \frac{2S}{R_o} \tag{9}$$

We can combine both thermodynamic equilibrium (Eq. (8)) and mechanical equilibrium (Eq. (9)) to get the desired correlation. Substituting Eq. (9) into Eq. (8), and after reorganizing we get,

$$R_0 = \frac{2S}{P_{sat}(T_\infty) \exp\{v_l[P_{nuc} - P_{sat}(T_\infty)]/R_v T_\infty\} - P_{nuc}} \tag{10}$$

Fig. 4 plots Eq. (10) for the different surface tension of gelatin concentration. Data from the water-vapor saturation table is used for the other parameters at temperature, $T_\infty = 20^\circ C$. Kang et al. (2018) reported that the mean tensile pressure for water was $120 kPa$, which corresponds to a critical radius of $1.2\ \mu m$ (Fig. 4). Even for the same critical radius to cavitate in the gel system without considering the elastic contribution, the tensile nucleation pressure would be higher due to increasing surface energy ($W_i$) to overcome (horizontal dashed line in Fig. 4). However, the pore size within the RFN of the gel system varies with the concentration of gel, and the effect of the microstructure on the critical nuclei radius will be discussed in subsequent sections.

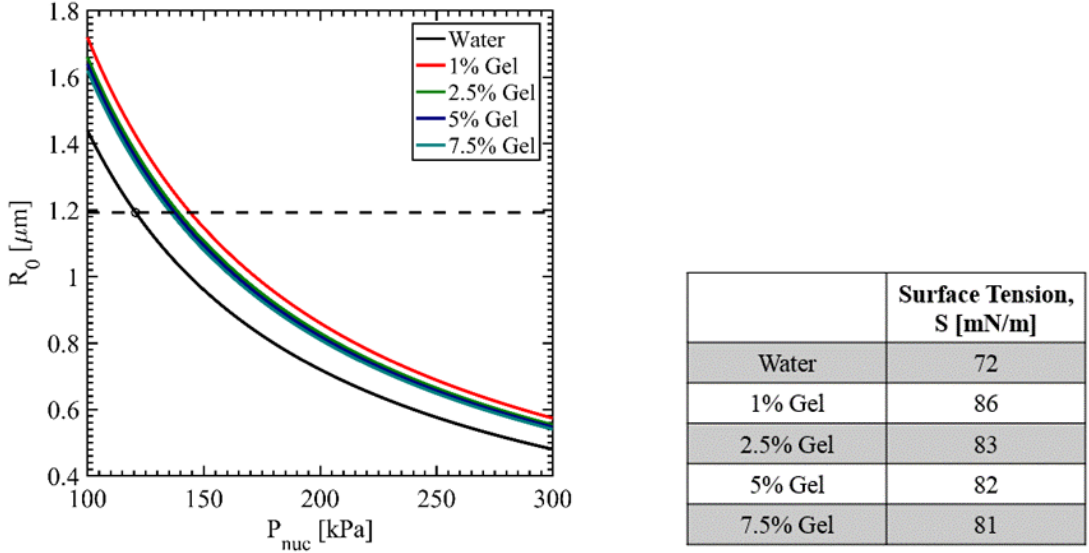

Fig. 4. Critical bubble radius as a function of the nucleation pressure from Eq. (10). The surface tension value is taken from Ref. [4] for the gelatin gel.

*2.2 Extra Tensile Pressure*

Considering the bubble only contains saturated vapor and no non-condensable gas, the mechanical equilibrium at forming a bubble in a pure liquid is given in Eq. (9). Where we assumed that $R_o$ is the *reference configuration* with the stress-free surrounding medium, hence there is no elastic contribution in Eq. (9). However, when a bubble starts to grow, strain energy is stored in the network. At the *current configuration,* the pseudo-mechanical equilibrium is established at the onset of cavitation in a gel as pressure further drops from $P_{nuc}$ to $P_T$,

$$P_v\left(\frac{R_o}{R}\right)^3 = \Delta P_T + \frac{2S}{R} + P_{RFN} \qquad (11)$$

Where, $P_{RFN}$ is the pressure contribution from the strain energy density stored in the network until the bubble stretch ratio ($\lambda_B = R/R_o$) reaches a critical value. In the above equation, we have considered vapor as an ideal gas with isothermal polytropic expansion ($k$=1). Therefore, the left-hand side represents the bubble pressure change as it grows. In terms of the bubble stretch ratio, Eq. (11) can be written for the extra tensile pressure as,

$$\Delta P_T = P_v \left(\frac{1}{\lambda_B}\right)^3 - \frac{2S}{R_o \lambda_B} - P_{RFN} \tag{12}$$

In this manuscript, we have adopted two approaches to study the contribution of the extra tensile pressure. The *strain-energy based failure criteria* and *fracture-based failure criteria* have been developed based on the random fiber network and critical energy release rate, respectively. The *strain-energy-based failure criteria* require a detailed study of the gel microstructure and fibers' material properties, which is discussed in the next section.

### 3. Random Fiber Network of the Gelatin Gel

Marmorat et al. (2016) used the cryo-SEM imaging technique to observe the gelatin supramolecular structure. They showed that gelatin fibers are connected via the triple-helical gelatin strand at the crosslink. However, in low crosslink density, the distance between the crosslinks is large to allow the strands' natural tendency to recoil into fibrils (Fig. 5). Those gelatin fibrils showed a well-known banded pattern with a periodicity of 64 $nm$ similar to the collagen fibril [73]. The theory of determining the mesh size based on the rubber elasticity [74] underestimated the mesh size observed by Marmorat et al. (2016). MacKintosh et al. (1995) proposed that the "semi-flexible" network shows increased rigidity due to its secondary structure between the crosslinks and must be taken into consideration to determine the mesh size [75]. Considering the network consists of effective spring where effective spring constant is $k_{ef}$, then the small strain shear modulus is related to the mesh size as [67],

$$\xi_0 = \frac{k_{ef}}{G} \tag{13}$$

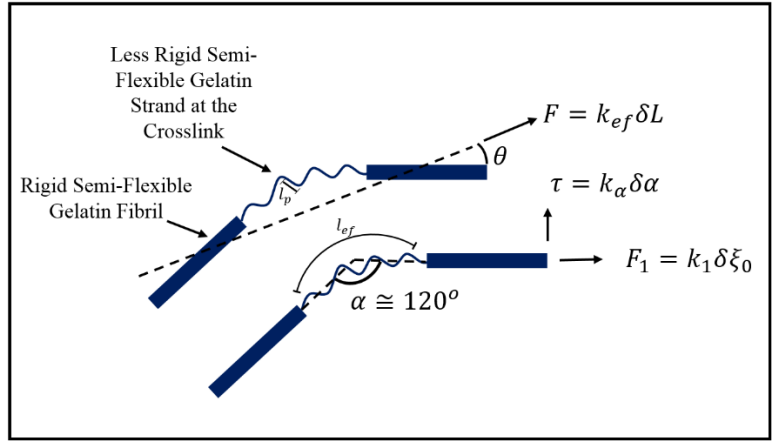

Fig. 5. Typical crosslink of the gelatin fiber network. Individual fibers (think blue line) are crosslinked via the gelatin strand (thin blue line).

For small strain, the rigidity arises from the straightening of $\alpha$ angle of the gelatin strand at the crosslink. However, Gelatin fibril will deform as well for large strain, and the large deformation of the fibril will be considered in later sections for the bubble growth. From Fig. 5, for any force ($F$) applied to the crossling with effective spring constant ($k_{ef}$) will have two components; i) $F_1$ will stretch the gelatin strand with stretching spring constant $k_1$, and ii) torque $\tau$ will act to change angle $\alpha$ with angular deformation constant $k_\alpha$. Treloar (1960) calculated the $k_{ef}$ for polymeric chain and will be adopted here [76],

$$\frac{1}{k_{ef}} = \frac{\delta L}{F} = n\left[\frac{cos^2(\alpha/2)}{k_1} + \frac{l_p^2 sin^2(\alpha/2)}{k_\alpha}\right] \tag{14}$$

In the above equation, $n = l_{ef}/l_p$, is the amount of zigzag, which is defined as the ratio of the effective length of the gelatin strand ($l_{ef}$) at the crosslink and the persistence length of the gelatin strand ($l_p$). The cryo-SEM observation showed that the statistical average crosslink angle is $\langle\alpha\rangle \cong 120^o$, hence $\langle sin^2\left(\frac{\alpha}{2}\right)\rangle = 0.75$ [67]. Considering crosslink deformation dominated by $k_\alpha$ for small strain and for an inextensible gelatin strand ($k_1 \gg k_\alpha$), we can drop the first term in Eq. (14),

$$k_{ef} = \frac{k_\alpha}{nl_p^2 \langle sin^2\left(\frac{\alpha}{2}\right)\rangle} \tag{15}$$

From Treloar (1960) and Marmorat (2016), the angular deformation constant is related to the persistence length and length of the bond along the polymer backbone ($l_0$) as $k_\alpha = k_B T_o l_p / l_0$. Where $k_B$ is the Boltzmann constant. Therefore, the effective spring constant is,

$$k_{ef} = \frac{4 k_B T_o}{l_0 l_{ef} \langle sin^2\left(\frac{\alpha}{2}\right)\rangle} \tag{16}$$

There exist two limiting cases for the effective length ($l_{ef}$) of the gelatin strand (Fig. 5). For high-density crosslink, the gelatin strand will not be able to recoil to form a superstructure. Therefore, fibers will be made of gelatin strand only, and the effective length will be, $l_{ef} \to \xi_0/2$. On the other hand, for low-density crosslinks, fibers will have enough strands to recoil, and gelatin fibril with superstructure will form. Hence, the minimum gelatin strand effective length at the crosslink, which straightens in the small strain, will be reduced. In that case, since the theoretical minimum length would be two persistence length to form the crosslink bend, the effective length would be, $l_{ef} \to 2l_p$. From Eq. (13) and Eq. (16),

$$\xi_0 = \sqrt{\frac{8 k_B T_o}{G l_0 \langle sin^2\left(\frac{\alpha}{2}\right)\rangle}} \qquad for\ l_{ef} \to \xi_0/2 \tag{17}$$

$$\xi_0 = \frac{2 k_B T_o}{G l_0 l_p \langle sin^2\left(\frac{\alpha}{2}\right)\rangle} \qquad for\ l_{ef} \to 2l_p \tag{18}$$

Eq. (17) and Eq. (18) is the theoretical minimum and maximum limit of the fiber length, respectively [67]. The small strain shear modulus is measured for 5, 7, and 14% [w/v] of Knox gelatin gels using a piezoelectric cantilever measurement technique and reported in Fig. 6a from the ref. [77]. The black dashed line shows a linear fit with $R^2 = 0.99$. The shear modulus value is used in Eq. (17) and Eq. (18), and the mesh size is plotted in Fig. 6b. The arithmetic mean of one C-C bond and two C-N bonds is used as the bond length along the polymer backbone ($l_0 = 1.4\ Å$) and persistence length ($l_p = 2.7\ nm$) for gelatin strand is used, respectively from the ref. [78] and [79]. The theory is validated by observing the SEM image of 3 different concentrations of gelatin gels. We have followed the procedure described in ref [4] for gelatin sample preparation. The detail of the sample preparation and the SEM imaging is given in the supplementary document. The SEM images are given in Fig. 6c, and the measurement of the mesh sizes of 3, 5, and 7% [w/v] gelatin gels are plotted in Fig. 6b and compared with Eq. (17) and Eq. (18). It seems the limiting case for Eq. (18) is valid, which means gelatin fibrils with the supramolecular structure are formed, and the gelatin strands connect crosslinks.

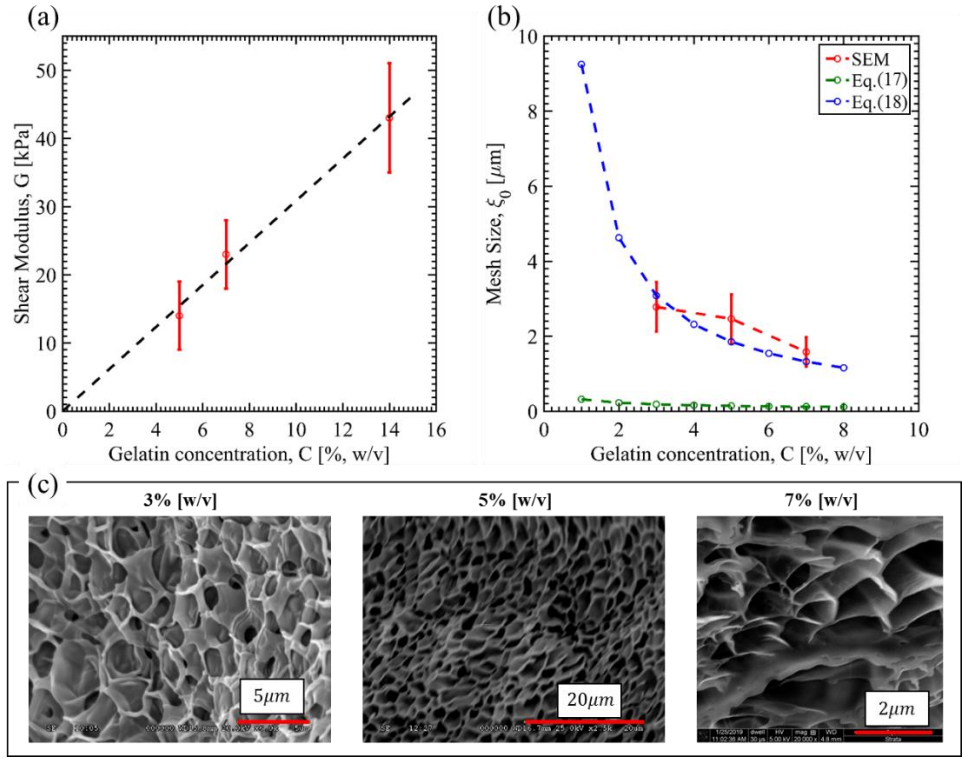

Fig. 6. (a) Shear modulus, (b) fiber length observed from the SEM images and theoretical model (Eq. (17) and(18)) fit, and (c) SEM images of different gel concentrations. The vertical red error bar indicates the standard deviation.

The pose size is then measured using the Diameterj plugin with Fiji (Imagej2) software [80], [81]. The SEM image is first converted to an 8-bit binary image and then segmented using 16 different algorithms provided by the Diameterj. Each image is then analyzed, and the mean pore size is measured. Fig. 7 shows an 8-bit SEM image, segmented image, and pose size measurement for 3% gel as an example. The detailed procedure is described in the supplemental document.

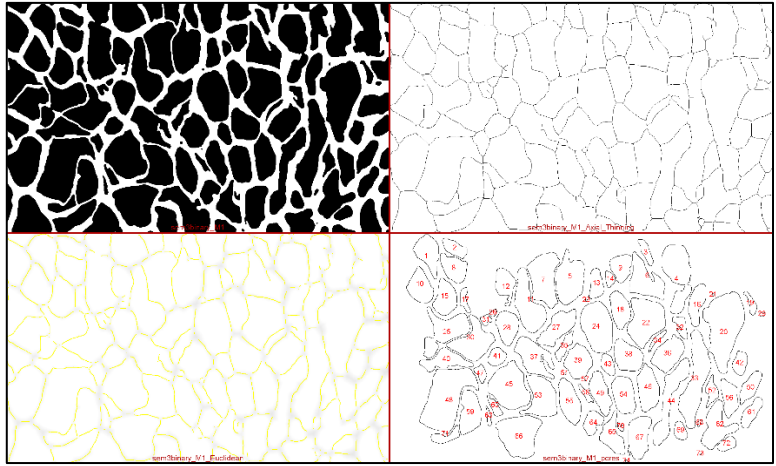

Fig. 7. 8-bit image of the SEM image of 3% gel (top left). Segmentation is done in two steps (top right and bottom left). Pore count and size measurement are shown (bottom right).

Fig. 8 shows the pore size ($A_p$), and since ImageJ fits the pore size in an ellipse, we have shown the minor axis length ($L_{MA}$) of the pores in Fig. 9. As we have postulated in the previous section, a bubble nucleus of the critical radius ($R_0$) must be accommodated within a pore to cavitate under the tensile nucleation pressure ($P_{nuc}$). The maximum size of a sphere that can fit in an ellipse must have a radius that is half of the minor axis length. Therefore, in Eq. (10), we will use $R_0 \to L_{MA}/2$ for gel system to compute $P_{nuc}$.

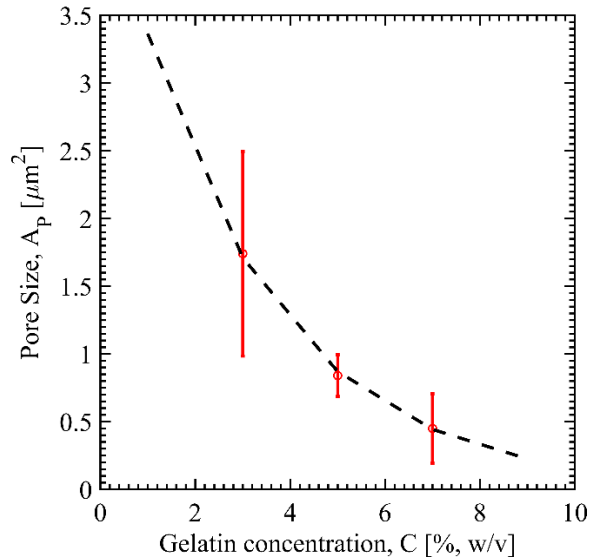

Fig. 8. Pore area size was measured by observing the SEM image of different gel concentrations. (Vertical red error bar indicates the standard deviation)

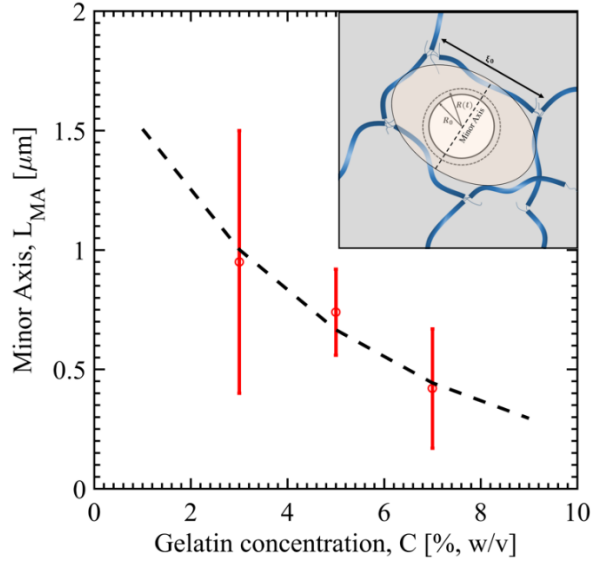

Fig. 9. Minor axis length of the pore area. (Vertical red error bar indicates the standard deviation)

## 4. Strain Energy-Based Failure Criteria

Since $l_{ef} \to 2l_p \ll \xi_0$, we will assume that the crosslink to crosslink fiber is consists of the gelatin fibril only. From now on, we will use gelatin fibril length and mesh size interchangeably. To construct the gelatin network, we will replace the contribution of the gelatin strand with torsional spring at the crosslink with effective torsional spring constant, $K_t$ (Fig. 11). Since gelatin triple-helical strands self-assemble into the secondary supramolecular structure as collagen fibril, which is modeled as semi-flexible fiber, we can model each gelatin fibril as such [44]. We used the word "fibril" for gelatin fiber to be consistent with the definition of the hierarchical structure of collagen micro-fibril, fibril, and fiber [82]. A semi-flexible fibril is defined such that its thermal persistence length is comparable to the fibril length, $L_P/\xi_0 \sim 1$. The persistence length of the gelatin fibril can be defined as the ratio between the bending rigidity and thermal energy, $L_P = E_f I_f / k_B T_o$, where $E_f$ is the fibril Young's modulus and $I_f = \pi d_f^4/64$ is the moment of inertia of the fibril, respectively. The material and geometric properties we need to compute the strain energy of the network are the fibril Young's modulus ($E_f$), crosslink torsional spring constant ($K_t$), fibril length ($\xi_0$), and diameter of the fibril ($d_f$). Considering the gelatin fibril as elastic beam which resists stretching and bending, and crosslink as torsional spring, the strain energy of the network for any given configuration is defined as [44],

$$U = U_S + U_B + U_T \qquad (19a)$$

$$= \sum_{i=1}^{N_s}\left(\int_0^{\Delta\xi_0} F_f d\Delta\xi_0\right)_i + \sum_{i=1}^{N_b}\left(\int_0^{\xi_0} \frac{M^2}{2E_f I_f} dx_f\right)_i + \sum_{j=1}^{N_t}\left(K_t \frac{\Delta\theta_t^2}{2}\right)_j \tag{19b}$$

In Eq. (19b), $N_s$ and $N_b$ are the number of fibrils that contribute to the stretching strain energy ($U_S$) and bending strain energy ($U_B$), respectively. Crosslink torsional strain energy ($U_T$) is due to the $N_t$ number of crosslinks having rotational angles $\theta_t$. The difference between the current and reference configuration is indicated by $\Delta$, and $A_f = \pi d_f^2/4$ is the cross-sectional area of the fibrils. The stretching force is defined as, $F_f = E_f A_f (\exp(B_f \varepsilon_f) - 1)/B_f$, where $B_f$ is a material parameter, and the Green strain ($\varepsilon_f$) is calculated using the fibril stretching ratio as, $\varepsilon_f = (\lambda_f^2 - 1)/2$ [83]. $M$ is the non-uniform (fibril length-wise) transverse moment on the fibril due to bending ($x_f$ is the fibril local longitudinal coordinate defined in Fig. 16).

Incorporating the bubble growth ratio ($\lambda_B$) into Eq. (19), we can compute the strain energy of the network. Ultimately, we will assume that the network failure strain energy is the work done by $P_{RFN}$. In the next sub-section, we will adopt a unit cell model to find fibril material and geometric properties. Then we will incorporate the bubble growth into Eq. (19).

*4.1. Unit Cell Model and Fibril Properties*

In the unit cell model, we will fit the macroscale stress-strain data to the unit cell's microstructural deformation. For large strain, gelatin gel is proposed to behave like incompressible hyperelastic Ogden material [69]. The strain energy density function for the first order Ogden material in terms of principal stretches of gel ($\lambda_{Gi}$) in three coordinate directions ($i = 1,2,3$) is,

$$W_G = \frac{\mu_G}{\beta}(\lambda_{G1}^\beta + \lambda_{G2}^\beta + \lambda_{G3}^\beta - 3) \tag{20}$$

Where $\mu_G$ and $\beta$ are material properties. Particularly $\beta$ is the strain hardening parameter, and for gelatin, $\beta$ is reported from ref [69] (plotted in Fig. 10). The small strain shear modulus is defined as, $G = \mu_G \beta/2$, and given in Fig. 6a. For uniaxial tension, the deformation gradient tensor $\boldsymbol{F_G}$ is,

$$\boldsymbol{F_G} = \begin{bmatrix} \lambda_{G1} & 0 & 0 \\ 0 & \lambda_{G2} & 0 \\ 0 & 0 & \lambda_{G3} \end{bmatrix} \tag{21}$$

The incompressibility condition is met when $\det(\boldsymbol{F_G}) = \lambda_{G1}\lambda_{G2}\lambda_{G3} = 1$. For uniaxial tension in direction 1, let $\lambda_{G1} = \lambda_G$ and $\lambda_{G2} = \lambda_{G3} = \lambda_G^*$, then from incompressibility condition, $\lambda_G^* = 1/\sqrt{\lambda_G}$. The three principal values of the Cauchy stress ($\boldsymbol{T_G}$) is defined as,

$$T_{Gi} = -\tilde{p} + \lambda_{Gi}\frac{\partial W_G}{\partial \lambda_{Gi}} \tag{22}$$

Where $\tilde{p}$ is the pseudo-pressure term that is determined for uniaxial tension by setting, $T_{G2} = T_{G3} = 0$. The experimental data reported in ref [69] is for the first Piola-Kirchhoff (nominal) stress, which is defined as, $\boldsymbol{P_G} = \det(\boldsymbol{F_G})\boldsymbol{T_G}(\boldsymbol{F_G}^T)^{-1}$. Therefore, the first Piola-Kirchhoff stress in direction 1 is (plotted in Fig. 10),

$$P_{G1} = \frac{2G}{\beta}\left(\lambda_G^{(\beta-1)} - \lambda_G^{(-\frac{\beta}{2}-1)}\right) \tag{23}$$

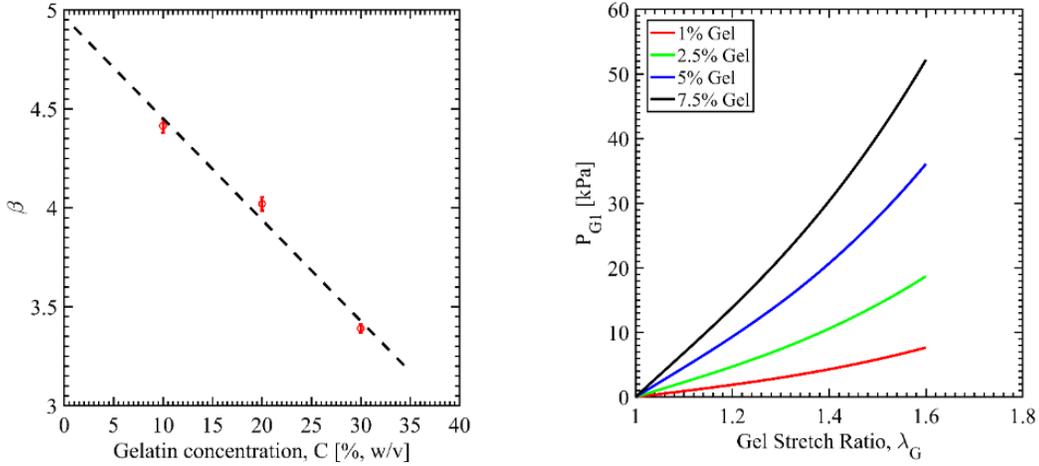

Fig. 10. Strain hardening parameter ($\beta$) for gelatin from uniaxial testing (left) [69]. The First Piola-Kirchhoff stress of gelatin gel as a function of the gel stretch (Eq. (23)) (right).

The three-dimensional microstructure of the gel system is represented by the unit cell in Fig. 11a. The proposed unit cell assumes isotropic microstructure, and the angles are set to be $\psi = 45°$ and $\phi = 35.26°$ for the unit cell to be symmetric in all principal coordinate directions. The other two geometric properties are fibril diameter ($d_f$) and length ($\xi_0$). The initial dimensions of the unit cell in three coordinate directions are,

$$L_{1,0} = 2\xi_0(1 + cos\phi cos\psi) \qquad (24a)$$

$$L_{2,0} = 2\xi_0(1 + sin\phi) \qquad (24b)$$

$$L_{3,0} = 2\xi_0(1 + cos\phi sin\psi) \qquad (24c)$$

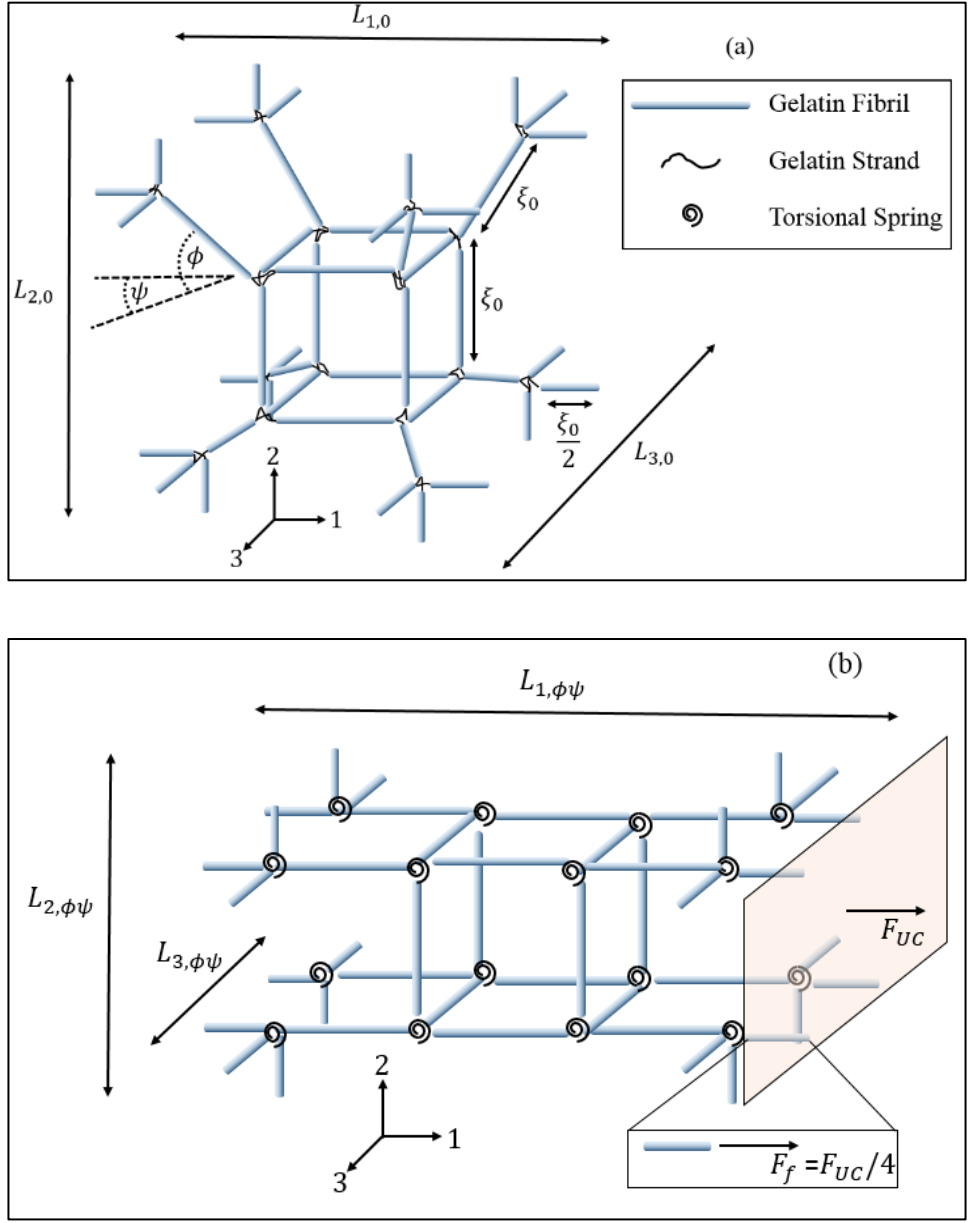

Fig. 11. (a) A representative unit cell of the random fiber network of gelatin gel. (b) Uniaxial stretching until fibers align in the stretch direction due to the crosslink rotation.

Previously, we have assumed that the small strain is due to the strengthening of the crosslink angles. Therefore, oblique fibrils are aligned parallel to direction 1 before fibril level stretch ($\lambda_f$) imposed on the fibrils (Fig. 11b). The dimension of the unit cell when fibril alignment occurs are,

$$L_{1,\phi\psi} = 4\xi_0 \tag{25a}$$

$$L_{2,\phi\psi} = 2\xi_0 \tag{25b}$$

$$L_{3,\phi\psi} = 2\xi_0 \tag{25c}$$

The gel stretch is defined as,

$$\lambda_G = \frac{L_1}{L_{1,0}} \tag{26}$$

In the current configuration, the dimension of the unit cell in direction 1 is $L_1$. The gel stretch at the fibril alignment is,

$$\lambda_{G,align} = \frac{L_{1,\phi\psi}}{L_{1,0}} \tag{27}$$

Since fibril stretch occurs after fibril alignment, the fibril stretch is defined as,

$$\lambda_f = \frac{L_1}{L_{1,\phi\psi}} = \frac{\lambda_G}{\lambda_{G,align}} \qquad for\ \lambda_G \geq \lambda_{G,align} \tag{28}$$

The first Piola-Kirchhoff stress is defined as the force per initial unit area. Therefore, the force on the unit cell ($F_{UC}$) is,

$$P_{G1} = \frac{F_{UC}}{L_{2,0}L_{3,0}} \tag{29}$$

From Fig. 11b, the force on the fibril ($F_f$) is one-fourth of the force on the unit cell, $F_f = F_{UC}/4$. From Eq. (19b), we can decompose the stored strain energy of the RFN as crosslink deformation and fiber stretching (Fig. 12),

$$W_G(\lambda_G)V_U = \begin{cases} U_T = \sum_{j=1}^{N_t=16} K_t \frac{\Delta\theta_t^2}{2} & for\ \lambda_G \leq \lambda_{G,align} \quad (30a) \\ U_S = \sum_{i=1}^{N_s=4} \left(\int_0^{\Delta\xi_0} F_f d\Delta\xi_0\right)_i - U_T(\lambda_{G,align}) & for\ \lambda_G > \lambda_{G,align} \quad (30b) \end{cases}$$

Where unit cell volume is $V_U = L_{1,0}L_{2,0}L_{3,0}$. Due to the symmetry, only one oblique fibril is shown in Fig. 12. The strain energy in Eq. (30a) is due to the rotation of the 16 crosslinks ($N_t = 16$) until $\lambda_{G,align}$, followed by four fibrils ($N_s = 4$) stretching until gel failure stretch, $\lambda_G^u$ (Fig. 12). $K_t$ and $E_f$, $d_f$ and $B_f$ will be computed from Eq. (30a) and (30b) using nonlinear regression analysis, respectively. 60% strain ($\lambda_G^u = 1.6$) is taken to be the failure criterion for the gelatin gels reported in ref. [69]. A similar failure strain is reported for collagen gels as well [44], [84]. Therefore, the fibril level stretch is computed from Eq. (28), which is $\lambda_f = 1.26$. Collagen fibril yield strain and ultimate failure strain are reported to vary between 10-32% and 35-45% strain, respectively [85]. Baumberger et al. (2006) observed that the fiber network failed by crosslink disentanglement at a low deformation rate and suggested that a higher strain rate network fails due to individual fiber scission [86]. Therefore, in the uniaxial testing at the quasi-static stretching, we can conclude that gel fails via crosslink failure, and fiber does not attain its failure strain. However, bubble growth imposes high strain rate deformation in the surrounding medium, and we will assume that fiber failure strain (35-45%) is the primary failure mechanism for bubble growth.

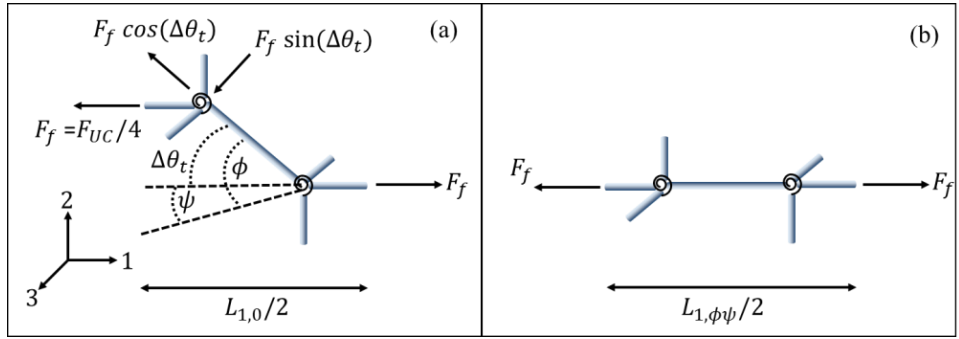

Fig. 12. (a) one oblique fiber is shown at the beginning of the uniaxial tension. The strain energy stored until the fibers align in the stretching direction is due to the crosslink's rotational strain energy. (b) fibers alignment is completed, and further stretching of the fiber stores energy due to stretching only.

There are several studies where collagen fiber properties are measured by uniaxial stress-strain testing. However, we have not found any material property for gelatin fibril to the best of our knowledge. Therefore, the unit cell model is developed, which can be utilized to estimate the gelatin fibril properties as the model can relate gel level properties to fibril level properties in close form. Since gelatin fibril and collagen fibril are similar in their supramolecular structure, we can use collagen fibril properties as a guideline. Susilo et al. (2010) have summarized the literature on collagen properties in detail, and readers are referred to that for further reading [39]. We have found that the collagen fiber properties vary in a wide range of magnitude. One of the main reasons for this wide range is the hierarchical structure of collagen fiber (e.g., microfibril, fibril, and fiber). Diameter is one of the parameters that can distinguish the collagen fibril from the collagen fiber. Collagen fibril is said to have a diameter ranging from 20 nm to 400 nm, while collagen fiber may have a diameter greater than 400 nm [87]–[89]. Fig. 13 plotted the gelatin fibril diameter for different concentrations using the unit cell model proposed earlier. The diameter range varies from 31 to 58 nm and well within the discussed range given above. The decreasing trend of the diameter is because the crosslink density increases with increasing concentration; therefore, there are fewer gelatin strands to recoil to form thicker fibrils. Cryo-SEM observation of increasing crosslink density showed a similar trend [67].

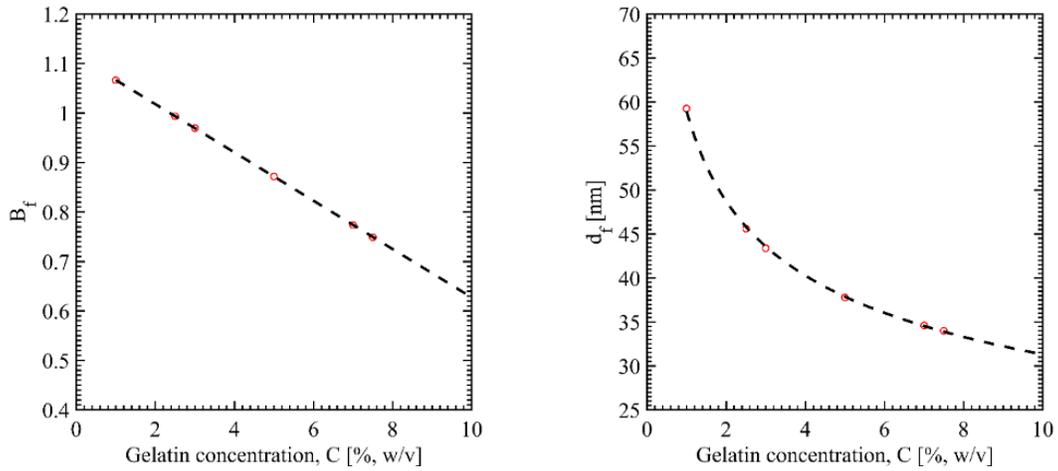

Fig. 13. Nonlinear material parameter (left) and diameter (right) of the fibers are shown for different gel concentrations.

The range mentioned above for collagen fibril diameter (20 to 400 nm) corresponds to fibril Young's modulus from 32 MPa to 11.5 GPa [90]–[92]. For collagen fiber ($d_f > 400 nm$) the modulus range is given to vary between 1.8 MPa to 1.2 GPa [39]. Fig. 14 plotted the Young's modulus of gelatin fibril for different concentration of gelatin gel by utilizing the unit cell model and gel level material properties as described above. The Young's modulus range is shown to be within 400 MPa to 1.15 GPa and falls within the collagen fibril modulus range. The fibril parameters will be used in the next sub-section to quantify the network strain energy due to bubble growth.

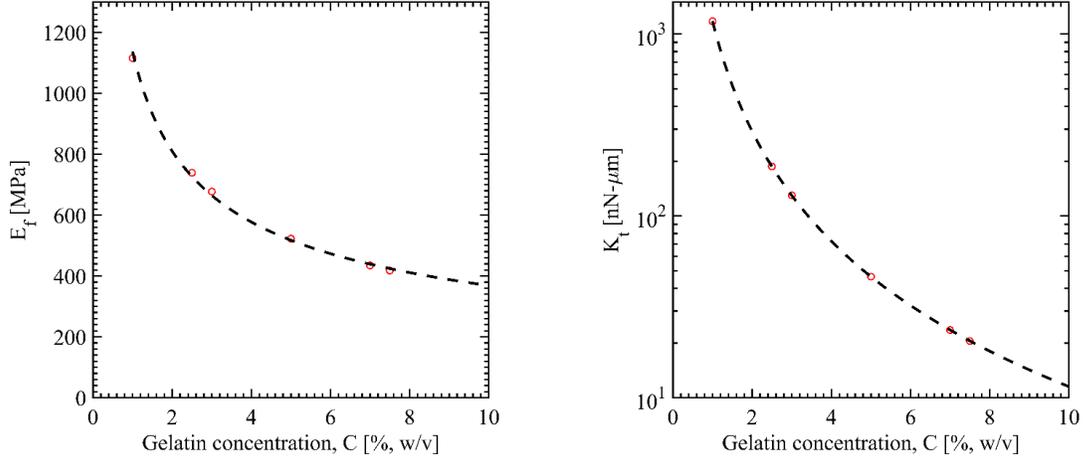

Fig. 14. Young's modulus (left) of the fiber and crosslink torsional constant (right) of the network, as shown for different gel concentrations.

*4.2. Network Strain Energy due to the Bubble Growth*

The network strain energy formulation shown in Eq. (19b) can be utilized if we can establish correlations between the bubble growth ratio ($\lambda_B$) and three strain energy parameters, i.e. $\Delta\xi_o$, $M$ and $\Delta\theta_t$. However, Eq. (19b) is the superposition of three different modes of deformations (e.g., stretching, bending, and torsion). We will correlate the bubble growth to each mode of deformation separately. In doing so, a unit cell of the gelatin network is shown in Fig. 15a, enclosing a bubble of radius $R_0$.

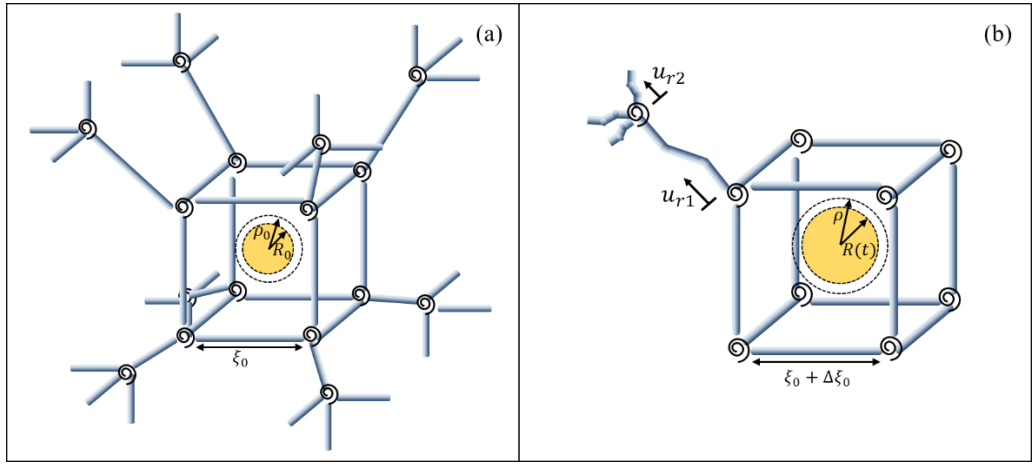

Fig. 15. (a) Bubble nucleus enclosed by the unit cell. (b) bubble growth causing fibril stretching (cubic part) and buckling (oblique fibril).

Considering in a strain-free reference configuration, we assume the bubble center lies at the origin of a spherical coordinate whose radius is $R_0$. In the current configuration at the time $t$, the bubble radius is $R(t)$. Any material point initially at $\mathbf{x_0} = \rho_0 \mathbf{e}_\rho$ will be at $\mathbf{x} = \rho \mathbf{e}_\rho$ in the current configuration. The deformation gradient tensor of the surrounding medium due to bubble growth is [5],

$$\mathbf{F_B} = \begin{bmatrix} \frac{\partial \rho}{\partial \rho_0} & 0 & 0 \\ 0 & \frac{\rho}{\rho_0} & 0 \\ 0 & 0 & \frac{\rho}{\rho_0} \end{bmatrix} \qquad (31)$$

The condition of incompressibility of the medium requires that $\det(\mathbf{F_B}) = 1$. Therefore, from Eq. (31), the radial stretch is, $\frac{\partial \rho}{\partial \rho_0} = (\frac{\rho_0}{\rho})^2$. Since $\rho_0 < \rho$ while bubble grows, the radial stretch decreases monotonically as a function of radial distance. Hence, in Fig. 15b, we have shown that all the oblique and end fibrils will experience buckling (only one corner is shown due to the ease of representation) except the fibril in the cubic portion of the unit cell. However, the buckling of semi-flexible filament is entropy dominant [26]. Since the strain energy formulation in Eq. (19b) assumes the negligible entropic contribution, we will only consider the cubic portion of the unit cell that will contribute to the strain energy, and buckled fibrils are omitted.

### 4.2.1. Fibril Stretching Strain Energy

In Fig. 15b, we have shown only the mode-I (stretching) deformation of the cubic cell. At the reference configuration, the initial volume of the liquid phase within the cubic cell is (from Fig. 15a),

$$V_0 = \xi_0^3 - \frac{4}{3}\pi R_0^3 \qquad (32)$$

Now, after an infinitesimal time increment, the bubble grows to $R(t)$ and displaces the surrounding liquid, which eventually interacts with the network and increases the fiber length to $\xi_0 + \Delta\xi_0$. The volume of the surrounding liquid within the displaced cubic cell is.

$$V = (\xi_0 + \Delta\xi_0)^3 - \frac{4}{3}\pi R^3 \qquad (33)$$

Assuming incompressibility of the liquid phase ($V_0 = V$), we can equate Eq. (32) and Eq. (33) to establish the correlation between the incremental fibril length, $\Delta\xi_0$ and the bubble extension ratio, $\lambda_B$ as,

$$\Delta\xi_0 = \frac{4}{9}\frac{\pi R_0^3}{\xi_0^2}(\lambda_B{}^3 - 1) \tag{34}$$

The failure criterion for Eq. (34) is set to be, $\Delta\xi_0{}^u = \xi_0(\lambda_f^u - 1)$. The fibril failure stretch ($\lambda_f^u$) is varied between 1.35 to 1.45, as discussed in the previous section.

*4.2.2. Fibril Bending and Crosslink Torsional Strain Energy*

Since the liquid phase is displaced radially outward in all directions, we can simplify the fibrils' bending shape and the crosslink rotation. From Fig. 16a, looking upon a cross-sectional view of the cubic cell cutting by the *A-A plane*, we can superimpose a circle of radius, $R_{AA} = \sqrt{3}\xi_0/2$ on the fibril deformation due to bending. Geometric similarity requires that on the *A-A plane,* the fractional area of the bubble growth (green shade) is related to the area ($A_2$) under the bent fibril (Fig. 16b). A deflection function ($y(x_f)$) is assumed for the fibril considering a simply supported beam with a constant distributed load per unit length ($q$) with local fibril coordinate system ($x_f, y_f$) is defined as well (Fig. 16c). The moment on the fibril is then defined as,

$$M = E_f I_f \frac{d^2 y}{dx_f^2} = \frac{q x_f}{2}(x_f - \xi_0) \tag{35}$$

The first derivative of the deflection function evaluated at either end of the fibril ($x_f = 0\ or\ \xi_0$) gives,

$$\Delta\theta_t = \frac{q\xi_0{}^3}{24 E_f I_f} \tag{36}$$

Since the area under the deflected fibril is,

$$A_2 = \int_0^{\xi_0} y\, dx_f = 0.207\pi R_o^2(\lambda_B^2 - 1) \tag{37}$$

Therefore, the distributed load on the fibril can be related to the bubble growth as,

$$q = \frac{24.84\pi}{\xi_o^5} E_f I_f R_o^2(\lambda_B^2 - 1) \tag{38}$$

Eq. (34), (35), (36), and (38) will be used with Eq. (19b) to compute the network strain energy due to bubble growth until fibrils failed at $\lambda_f^u$.

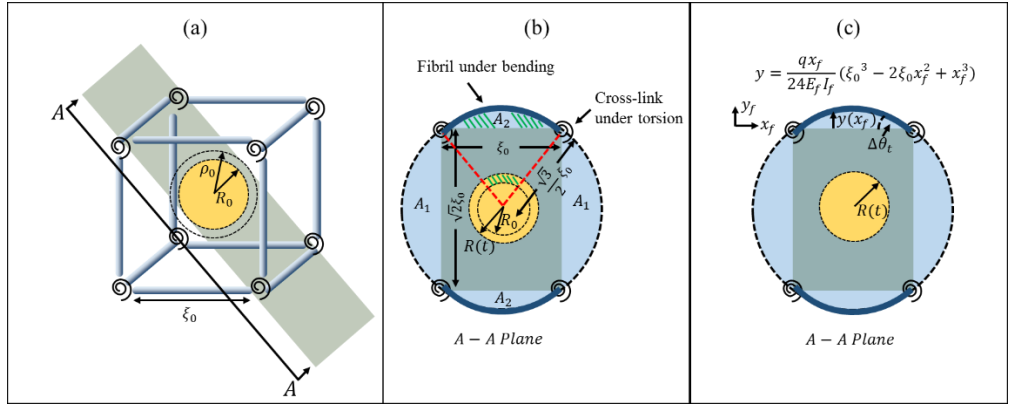

Fig. 16. (a) a diagonal plane (A-A) is shown on which fibril bending occurs due to the bubble growth. (b) and (c) projection on the A-A plane showing geometric relation of bubble growth area and fiber bending.

*4.3. Extra Tensile Pressure*

Now that we have formulated the strain energy stored in the network due to the bubble growth, we can define the pressure contribution from the strain energy per unit volume as,

$$P_{RFN} = \frac{U_{cric}}{\Delta V_{cric}} \tag{39}$$

In the above equation, $U_{cric} = U(\lambda_f^u)$ from Eq. (19). $\Delta V_{cric} = V_U - V_B^u$ is the difference between the volume of the unit cell and the volume of the growing bubble. The bubble volume is defined as,

$$V_B = \frac{4}{3}\pi R_o^3 (\lambda_B^3 - 1) \tag{40}$$

At the critical condition, $V_B^u = V_B(\lambda_B^u)$. The fibril failure stretch and bubble failure stretch are related through Eq. (34) as,

$$\lambda_f^u = 1 + \frac{4}{9}\frac{\pi R_o^3}{\xi_o^3}(\lambda_B^{u\,3} - 1) \tag{41}$$

Therefore, from Eq. (12), the extra tensile pressure is,

$$\Delta P_T = P_v \left(\frac{1}{\lambda_B^u}\right)^3 - \frac{2S}{R_o \lambda_B^u} - \frac{U_{cric}}{\Delta V_{cric}} \tag{42}$$

In the next section, we will develop the necessary formulation to quantify the extra tensile pressure based on the fracture theory.

## 5. Fracture Based Failure Criteria

In the previous section, we have formulated the extra tensile pressure criteria based on the failure of the RFN of the gel system. This section will use the well-known Griffith's criterion for gelatin fracture due to bubble growth, considering gel as the homogeneous hyperelastic Ogden material. We know that elastic materials can store energy when deformed and return to their reference configuration by spending that stored energy upon withdrawal of the loading. However, there is a limit on the stored energy beyond which fracture initiates and materials fail. Griffith's theorem states that a crack will propagate when surface energy is exceeded by the energy released due to new crack growth [93], [94]. The critical energy release rate ($G_c$) is the material property that is used as the criterion for the material to resist fracture. Wire cutting tests are done to estimate $G_c$ for gelatin gel and reported in Fig. 17 from ref. [69], [86], [95].

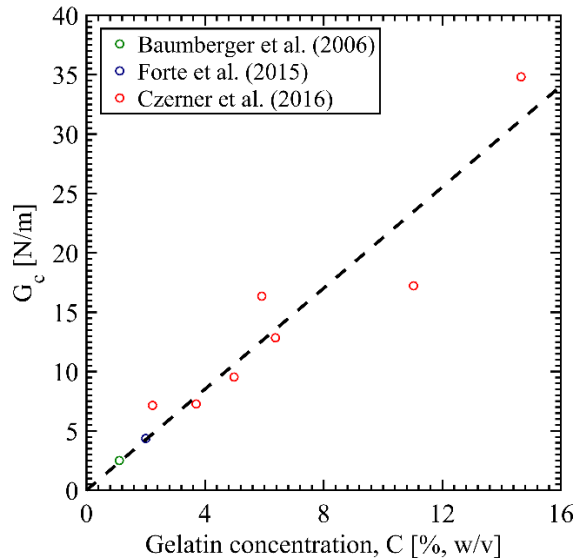

Fig. 17. Critical energy release rate as a function of the gelatin concentration.

Several authors have modified the Rayleigh-Plesset equation of bubble dynamics for viscoelastic materials. The elastic term is added for linear Hookean, nonlinear Neo-Hookean, and strain hardening Fung models [1], [5], [96]. They have provided the procedure in detail, and readers are referred to them for further study. However, in this manuscript, we will develop the elastic contribution for the Ogden material model.

From Eq. (31), the deformation gradient tensor of the surrounding medium in spherical coordinate direction ($i = \rho, \theta, \varphi$) is,

$$\boldsymbol{F_B} = \begin{bmatrix} \lambda_{B,\rho\rho} & 0 & 0 \\ 0 & \lambda_{B,\theta\theta} & 0 \\ 0 & 0 & \lambda_{B,\varphi\varphi} \end{bmatrix} = \begin{bmatrix} \dfrac{\partial \rho}{\partial \rho_0} & 0 & 0 \\ 0 & \dfrac{\rho}{\rho_0} & 0 \\ 0 & 0 & \dfrac{\rho}{\rho_0} \end{bmatrix} \tag{43}$$

The incompressibility condition is applicable here as well, which requires,

$$\frac{\partial \rho}{\partial \rho_0} = \left(\frac{\rho_0}{\rho}\right)^2 \tag{44}$$

Integrating the above equation and setting boundary condition at the bubble wall, $\rho = R(t)$ we get the reference coordinate in terms of the current coordinate of the material point,

$$\rho_0 = (\rho^3 - R(t)^3 + R_0^3)^{\frac{1}{3}} \tag{45}$$

From Eq. (22), the Cauchy stress tensor is,

$$\boldsymbol{T_G} = \begin{bmatrix} -\tilde{p} + \dfrac{2G}{\beta}\left(\dfrac{\rho_0}{\rho}\right)^{2\beta} & 0 & 0 \\ 0 & -\tilde{p} + \dfrac{2G}{\beta}\left(\dfrac{\rho}{\rho_0}\right)^{\beta} & 0 \\ 0 & 0 & -\tilde{p} + \dfrac{2G}{\beta}\left(\dfrac{\rho}{\rho_0}\right)^{\beta} \end{bmatrix} \tag{46}$$

In the above equation, we used $\lambda_{B,i}$ from Eq. (43) for the partial derivative of the strain energy density function. The pseudo-pressure term ($\tilde{p}$) in Eq. (22) and Eq. (46) is related to the hydrostatic pressure ($p$) as [1], [5],

$$p = -\frac{T_{G,\rho\rho} + T_{G,\theta\theta} + T_{G,\phi\phi}}{3}$$

$$= \tilde{p} - \frac{2G}{3\beta}\left[\left(\frac{\rho_0}{\rho}\right)^{2\beta} + 2\left(\frac{\rho}{\rho_0}\right)^{\beta}\right] \tag{47}$$

Finally, the elastic contribution to the bubble dynamics is found by integrating the momentum equation from the bubble wall, $\rho = R(t)$ to $\rho \to \infty$ [1], [5], [31], [96],

$$\Sigma = \int_{R(t)}^{\infty} -\left[\frac{2T_{G,\rho\rho} - T_{G,\theta\theta} - T_{G,\phi\phi}}{\rho}\right] d\rho$$

$$= \int_{R(t)}^{\infty} -\frac{4G}{\beta}\left[\frac{\rho_0^{2\beta}}{\rho^{2\beta+1}} - \frac{\rho^{\beta-1}}{\rho_0^\beta}\right]d\rho \tag{48}$$

Setting $\gamma = \rho/\rho_0$ and using Eq. (45), we get,

$$d\gamma = \left(\frac{1}{\rho_0} - \frac{\rho^3}{\rho_0^4}\right)d\rho \tag{49}$$

Considering, $\lambda_{B,\theta\theta} = \lambda_{B,\varphi\varphi} = \lambda_B = R/R_0$, Eq. (48) becomes,

$$\Sigma = \int_{\lambda_B}^{1} -\frac{4G}{\beta}\left[\gamma^{-(2\beta+1)}\left(\frac{1-\gamma^{3\beta}}{1-\gamma^3}\right)\right]d\gamma \tag{50}$$

The above equation is integrated numerically using the quadrature theorem. The stored strain energy due to the elastic contribution is [1], [15],

$$U_\Sigma = \int_{R_0}^{R} 4\pi\Sigma\rho^2\,d\rho \tag{51}$$

The energy release rate per unit crack area is defined as $\partial U_\Sigma/\partial A_c$, where $A_c = \pi R^2$ is the crack area. Applying Griffith's criterion of fracture on the energy release rate we get,

$$-\left(\frac{\partial U_\Sigma}{\partial A_c}\right) \geq 2R_0 f(\lambda_B) \tag{52}$$

Where $f(\lambda_B)$ is (using two dummy variables $(\zeta, and\ \eta)$),

$$f(\lambda_B) = \lambda_B^4 \frac{\partial}{\partial \zeta}\left(\zeta^{-3}\int_1^\zeta \Sigma\eta^2 d\eta\right)\bigg|_{\zeta=\lambda_B} \tag{53}$$

At the critical condition, bubble fracture stretch is defined as, $\lambda_B^u = R_u/R_0$. Therefore, Eq. (50) can be numerically integrated until the fracture stretch by setting $f(\lambda_B^u) = G_c/2R_0$ to find $\lambda_B^u$. We can then use Eq. (51) to find the total fracture energy at failure,

$$U_{\Sigma\,cric} = \int_{R_0}^{R_u} 4\pi\Sigma\rho^2\,d\rho \tag{54}$$

Therefore, the extra tensile pressure can be defined in terms of the total fracture energy per unit volume of the bubble growth as,

$$\Delta P_T = \frac{U_{\Sigma_{cric}}}{V_B^u} \tag{55}$$

## 6. Results and Discussion

### 6.1. Random Fiber Network

In developing the strain energy-based criterion for the onset of cavitation in soft materials, we have estimated the fibril and the network properties ($E_f, d_f, B_f, and\ K_t$) from the gel level properties ($G\ and\ \beta$) of the gelatin gels. The network's strain energy formulation considers fibrils as an elastic beam, and they store energy by stretching and bending. This elastic beam model is justified since the fibril's persistence length is much higher than the mesh size. Fig. 18 shows the fibril thermal persistence length and how it compares to the mesh size. Collagen fibers, having mesh size in the micrometer range ($\sim 2\mu m$) and thermal persistence length in the millimeter range ($\sim 10 mm$), are often modeled as the elastic beam [44], [45]. $L_P/\xi_o < 2/\pi^{3/2}$ and $L_P/\xi_o \gg 1$ correspond to "flexible" and "rigid rod" type fibers, respectively [26]. A flexible fiber's elastic response is due to the decrease in entropic conformation, and rigid rod-type fibers are modeled as beam theory. In between, there lies the definition of the "semi-flexible" fibers ($L_P/\xi_o \sim 1$) which shows elastic response by both stretching and bending. At low concentration and lower molecular weight, a semi-flexible fiber network shows *nonaffine* deformation, and fiber response is mainly due to bending. As concentration increases, the network tends to show more *affine* deformation dominated by fiber stretching [49]. In the crosslinked network, the geometric persistence length is comparable to the mesh size, and fibers are considered semi-flexible.

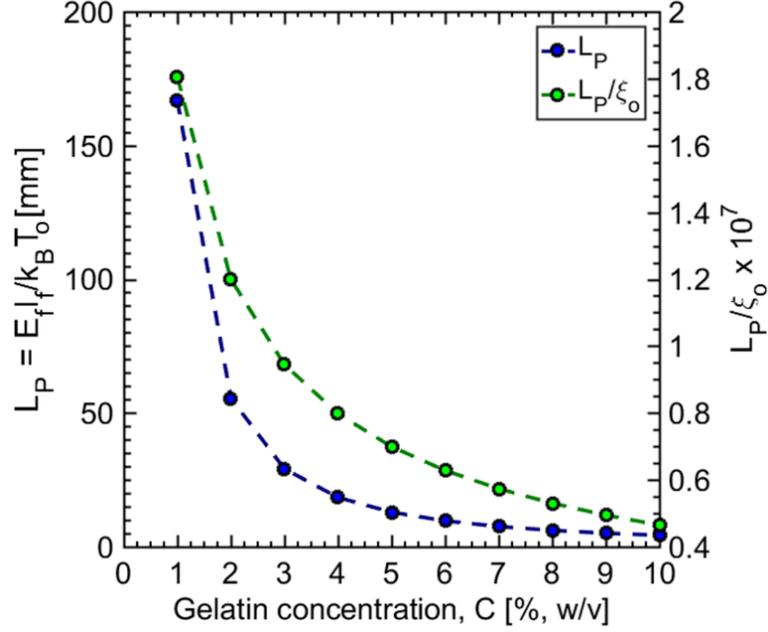

Fig. 18. The persistence length of the gelatin fibers is plotted on the left axis, based on the geometric ($d_f$) and mechanical ($E_f$) properties reported in Fig. 13 and Fig. 14, respectively. The ratio between the persistence length and mesh size (Fig. 6b) is plotted on the right axis. Both are shown as a function of the gel concentration.

Using Eq. (19), the network's strain energy for different gel concentrations is plotted in Fig. 19. To quantify the contribution of different modes of deformation, the critical values (see Eq. (39)) are reported for the initial bubble radius of $1.2 \mu m$ and fiber failure strain is taken to be 40% ($\lambda_f^u = 1.40$). The total critical strain energy ($U_{cric}$) is due to the stretching ($U_S$) of the fibers, since both bending energy of the fiber ($U_B$) and crosslink rotational energy ($U_T$) are few orders of magnitude lower than the stretching energy. Although the bending energy contribution increases slightly for low concentration, the network deformation is mainly affine and stretching dominant.

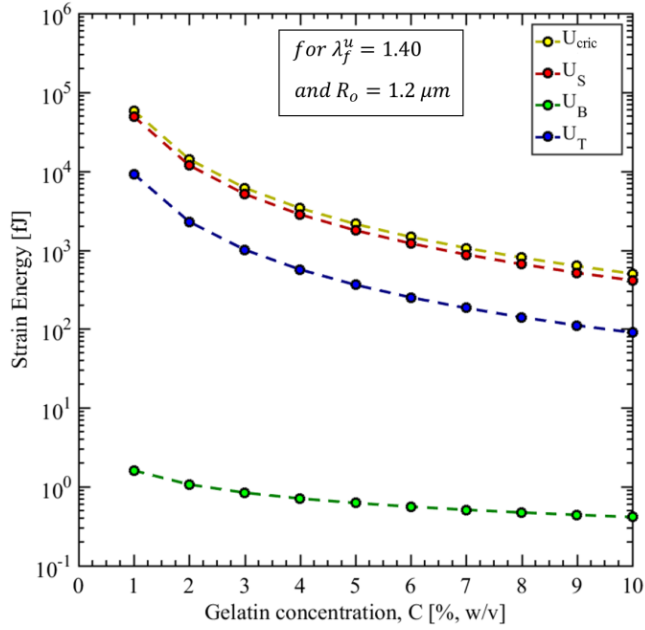

Fig. 19. The strain energy of the network due to the bubble growth from Eq. (19) corresponds to $R_o = 1.2\mu m$ and $\lambda_f^u = 1.40$.

*6.2. Critical Nucleation Pressure*

Fig. 20 plots the critical bubble size as a function of the gelatin gel concentration. The blue dotted line depicts the maximum bubble size that can be inscribed in the cubic portion of the unit cell model, which is half of the diagonal of the side face of the cube $(R_o = \xi_o/\sqrt{2})$. However, the network's SEM image observation indicates a much smaller *mean pore size* with a high standard deviation for the gelatin gels' low concentration (see Fig. 8 and 9). The maximum bubble size that can be formed within the network is then limited by the pore size (red dotted line) and defined as half of the pores' minor axis length $(R_o = L_{MA}/2)$. The nucleation formation energy in pure liquid corresponds to the critical nuclei size $(R_o = 1.2\ \mu m)$ from experimental observation [4], and shown as the green dotted line.

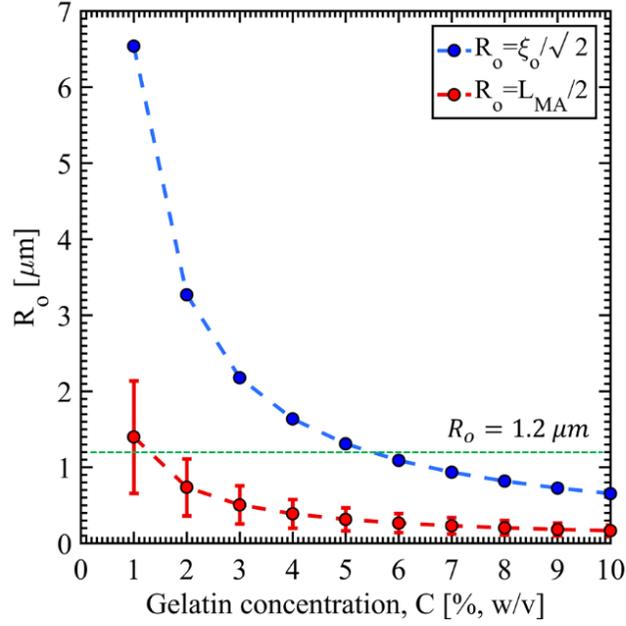

Fig. 20. Critical bubble radius based on the unit cell model and pore size from the SEM image.

A random fiber network, having a large distribution of pore size, may have enough large pores where nucleation formation energy will only depend on the surface energy $(2S/R)$. However, the surface tension $(S)$ value depends on various factors, such as the gel states (solution or gel), temperature, interface curvature, liquid inter-molecule affinity in the presence of the gelatin network [72], [97], [98]. In Fig. 21a, we have shown the surface tension of gelatin gel, both in solution and gel states, and compared it with water surface tension $(S_{water} = 72 mN/m)$ [4], [99], [100]. In the solution state, surface tension $(S_{sol})$ drops compared to the pure water while increases in the gel state $(S_{gel})$. The gelation process, however, starts by nonspecific hydrophobic interaction and the secondary supramolecular fibril forms [53]. As the network topology reaches rigidity percolation, more triple helical morphology starts to form. In the triple helical configuration, the nonpolar residues are arranged away from the water molecules. The polar molecules preferentially arrange outward of the tropocollagen. Therefore, as time passes during the gelation process the net attractive force between collagen and water molecules increases, thus the surface tension of the gel increases in the gel state [101]. However, gelation process is a chemically dynamic equilibrium state and never ends. The coexistence of the free-floating gelatin monomers ($\alpha$-chain) and the network is a typical condition in the gel system. Therefore, we have considered both surface tension (solution and gel) as the maximum and minimum limit for the nucleation. In Fig. 21b, we have plotted the nucleation pressure (Eq. (10)) for varying critical nucleation radius and represented as a function of the gel concentration. The vertical bars correspond to the maximum and minimum surface tension. While nucleation pressure for water is 120kPa for $R_o = 1.2\ \mu m$, it varies from 142 to 133 kPa for 1% to 10% of gelatin. Considering there exists enough large pore size in the network comparing to the critical bubble size that of water, we will assume this nucleation energy as the base case.

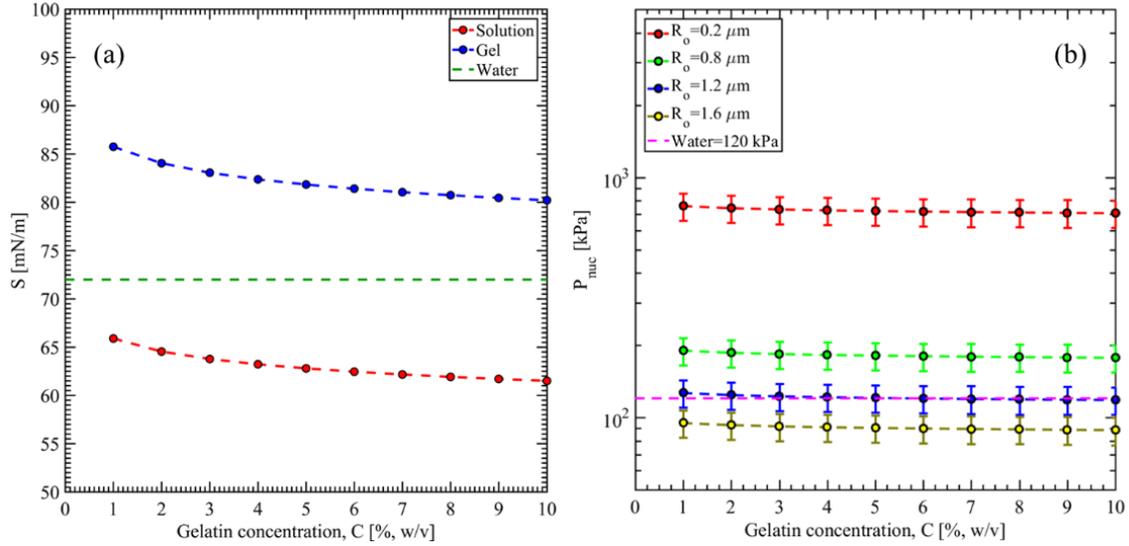

Fig. 21. (a) The surface tension of gelatin gel in solution and gel states. (b) Nucleation energy required for different bubble sizes and gelatin concentration compared to pure liquid (i.e., water).

*6.3. Critical Threshold Tensile Pressure*

In this section, we have summarized the extra tensile pressure ($\Delta P_T$) and the critical tensile pressure ($P_T$) on the onset of cavitation for the gelatin gels. Results from both the strain energy-based criteria (unit cell model) and the fracture-based criteria are presented. For the unit cell model, the critical condition is set to the fibril failure stretch ($\lambda_f^u$) and varied between 35 to 45% strain. Bubble failure stretch ($\lambda_B^u$) is used for the fracture-based model and set to $f(\lambda_B^u) = G_c/2R_0$. Two cases are considered for both models, i) critical bubble radius is kept fixed for all gel concentration, and ii) varied based on the unit cell maximum ($R_o = \xi_o/\sqrt{2}$), and pore size maximum ($R_o = L_{MA}/2$). Results are compared with the theoretical range ($\Delta P_T\{max, min\} = \{9G, 5G/2\}$) for the extra tensile pressure and experimental data for the critical tensile pressure from the work of Kang et al. (2018). The vertical bars are used for the unit cell results to indicate the maximum and minimum case corresponds to ($S_{gel}$ and $\lambda_f^u = 1.45$) and ($S_{sol}$ and $\lambda_f^u = 1.35$), respectively.

Fig. 22a shows the extra tensile pressure for case (i). Both the model predicts extra tensile pressure within the theoretical range for the given gel concentration. For low concentration, $\Delta P_T$ does not depend on the critical radius but varies in a wide range for higher gel concentration. $9G$ curve well predicts at the low concentration and overestimates as the concentration increases. $5G/2$ curve does not predict the extra tensile pressure at the low concentration as experimental observation suggests [14]. The unit cell model coincides with the theoretical maximum curve for the low concentration. As the concentration increases, the fiber length ($\xi_o$) decreases and the unit cell volume decrease as well. For the same bubble radius, the slope of the $\Delta P_T$ tends to reduce as the fiber failure strain is reached earlier, and less and less strain energy is required for the bubble to grow unconditionally. The gel critical energy release rate linearly increases with the gel concentration for the fracture model and shows an increasing trend. Fig. 22b plotted the extra tensile pressure for case (ii), where $R_o$ is a function of gel concentration. Two values of $R_o$ are considered as mentioned earlier. $R_o = L_{MA}/2$ is the lower limit of the maximum possible bubble based on the pore size distribution, and extra tensile pressure happens to coincide with the maximum limit given by $9G$ curve. As the maximum limit is based on smaller bubble nuclei with higher surface energy, $R_o = L_{MA}/2$ is more conservative and overestimates the critical tensile pressure. On the other hand, $R_o = \xi_o/\sqrt{2}$ is based on the maximum possible bubble size that can fit in the cubic portion of the unit cell. $R_o = \xi_o/\sqrt{2}$, when used in the unit cell model, predicts the most moderate values of the extra tensile pressure.

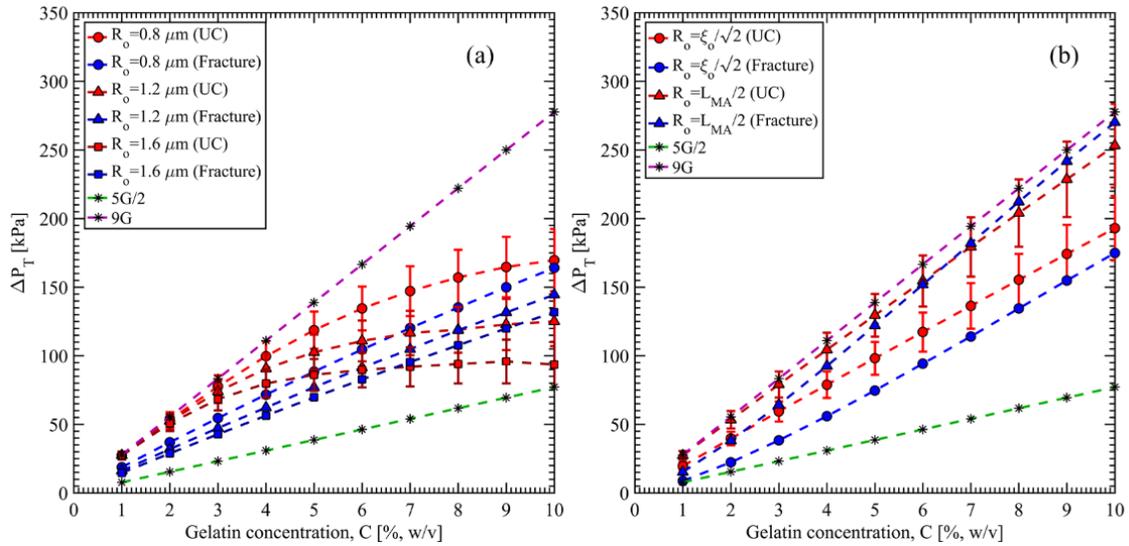

Fig. 22. Extra tensile pressure for (a) fixed critical radius for different gel concentrations, (b) varying critical radius. (UC=unit cell model, Fracture=fracture-based model)

Threshold tensile pressure is plotted and compared with the experimental data for the case (i) and (ii) in Fig. 23 and 24, respectively. Both unit cell and fracture-based model results are shown. For case (i), the best fit is provided by the $R_o = 1.2 \mu m$ and can predict the critical tensile pressure for a wide range of the gel concentration for both models. For case (ii), $R_o = \xi_o/\sqrt{2}$ can predict the critical tensile pressure well for the gel concentration from 3 to 7%. $R_o = L_{MA}/2$ overestimates the critical tensile pressure for both models, as discussed earlier.

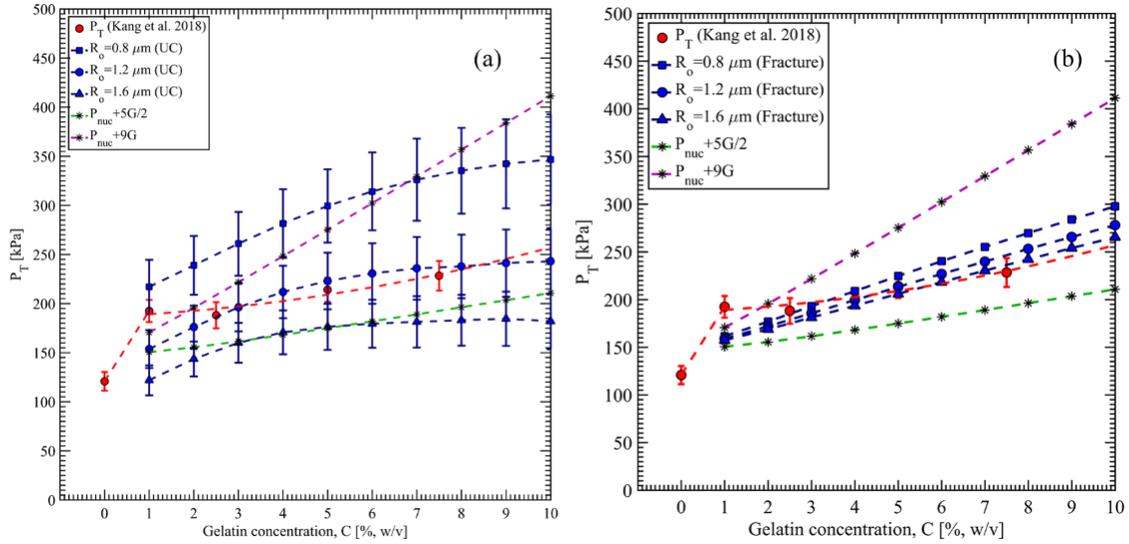

Fig. 23. Threshold tensile pressure for fixed critical bubble radius for different gel concentrations. (a) unit cell model and (b) the fracture model.

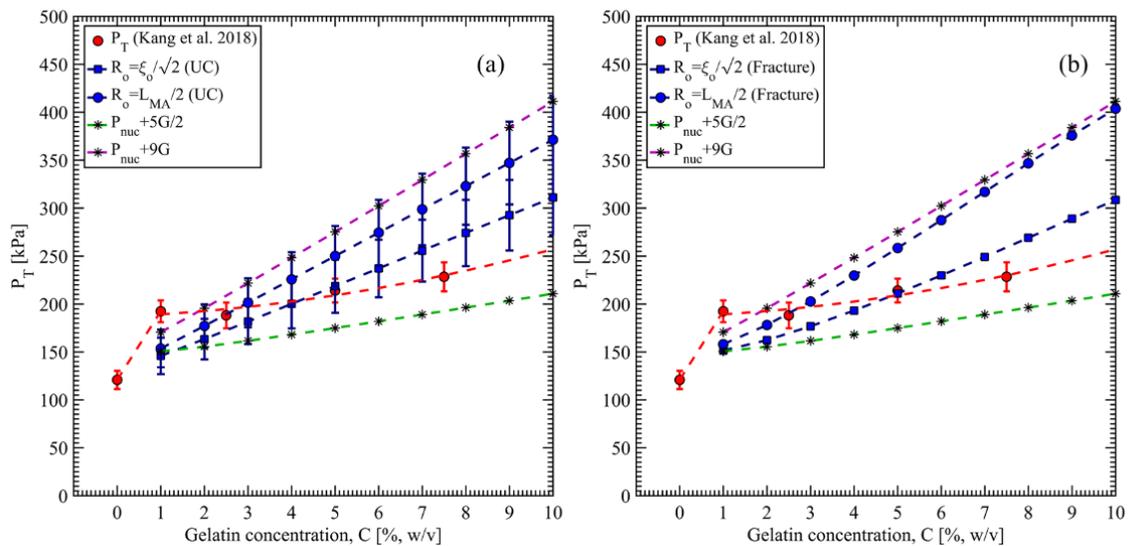

Fig. 24. Threshold tensile pressure for variable critical bubble radius for different gel concentrations. (a) unit cell model and (b) the fracture model.

## 7. Conclusion

In this manuscript, we have developed a theoretical framework to estimate the critical tensile pressure for cavitation in soft materials. Gelatin gel has been used for modeling purposes since it has various applications in the biomechanics field. Multifunctionality, less toxicity, and less biodegradability are the few reasons gelatin gels are being used for tissue engineering and tissue mechanics. In the first part of this manuscript, we have evaluated the gelatin network topology and then estimated the fiber properties by proposing a unit cell model. A bubble-network interaction is introduced, and strain energy-based failure criteria are then proposed. A fracture-based model is developed as well, and critical tensile pressure is evaluated for both failure criteria. As we have postulated, the nucleation pressure in gelatin is comparable to that of water. A large distribution of the pore size is the basis of this hypothesis. There exist enough large pores that can activate nucleation sites in the range of $\sim 1.2 \mu m$. The critical tensile pressure is well predicted by both unit cell and fractured based model for the critical bubble radius, $R_o \sim 1.2 \mu m$. However, both models underestimate the critical tensile pressure for 1% gel. Since the network rigidity percolation transition happens at the 1% gelatin, an affine network model is not adequate. Just after the network rigidity percolation, there exists a nonaffine domain. Nonaffine network elasticity is bending dominated and not considered in our proposed model. The nonaffinity is the measure of the heterogeneity of the deformation. Including the degree of the nonaffinity into the network-model, may improve its prediction for the low concentration gels. On the other side of the spectrum, we have a high concentration of the gelatin gel (~10%). As the thermal persistence length decreases with increasing concentration, the fibers are more entropic (flexible filament) than enthalpic. Cryo-SEM image observation showed that the fibers could not recoil and form secondary structures at the gelatin's higher crosslinked network. Since the crosslink density increases with increasing concentration, fibers need to be modeled as entropic. Several entropic fiber models are well discussed in the literature, such as the Gaussian, inextensible worm-like-chain (WLC) model, etc. We have not considered this approach since it is beyond the scope of this work.

**CRedit authorship contributions statement**

**Fuad Hasan:** Conceptualization, Methodology, Software, Validation, Formal analysis, Investigation, Writing - Original Draft. **KAH Al Mahmud:** Resources, Writing - Original Draft. **MD Ishak Khan:** Resources, Visualization. **Wonmo Kang:** Conceptualization, Validation. **Ashfaq Adnan:** Conceptualization, Methodology, Validation, Writing - Review & Editing, Funding acquisition.

**Declaration of competing interest**

The authors declare that they have no known competing financial interests or personal relationship that could have appeared to influence the work reported in this paper.

**Acknowledgements**

This work has been funded by the Computational Cellular Biology of Blast (C2B2) program through the Office of Naval Research (ONR) (Award # N00014-18-1-2082- Dr. Timothy Bentley, Program Manager).